\documentclass[twocolumn,pra,showpacs]{revtex4}

\usepackage{amsfonts}
\usepackage{amsmath}
\usepackage{graphicx}

\begin{document}

\title{Quantum regression theorem for non-Markovian Lindblad equations}
\author{Adri\'{a}n A. Budini$^{1,2}$}
\affiliation{$^{1}$Max Planck Institute for the Physics of Complex Systems, N{\"o}%
thnitzer Stra{\ss }e 38, 01187 Dresden, Germany\\
$^{2}$Instituto de Biocomputaci\'{o}n y F\'{\i}sica de Sistemas Complejos,
Universidad de Zaragoza, Corona de Arag\'{o}n 42, (50009) Zaragoza, Spain}
\altaffiliation{present address}
\date{\today}

\begin{abstract}
We find the conditions under which a quantum regression theorem can be
assumed valid for non-Markovian master equations consisting in Lindblad
superoperators with memory kernels. Our considerations are based on a
generalized Born-Markov approximation, which allows us to obtain our results
from an underlying Hamiltonian description. We demonstrate that a
non-Markovian quantum regression theorem can only be granted in a stationary
regime if the dynamics satisfies a quantum detailed balance condition. As an
example we study the correlations of a two level system embedded in a
complex structured reservoir and driven by an external coherent field.
\end{abstract}

\pacs{ 05.30.Ch, 03.65.Yz, 42.50.Lc, 03.65.Ta}
\maketitle

\section{Introduction}

In many areas of physics one is confronted with the description of small
quantum systems interacting with an uncontrollable environment. This
situation is well understood when the reduced system dynamics follows a
(completely positive) Markovian evolution \cite%
{alicki,nielsen,blum,carmichael,cohen,loudon}.

One of the cornerstone of the theory of Markovian open quantum systems is
the quantum regression theorem (QRT). This theorem, originally proposed by
Lax~\cite{lax}, allows to calculate multiple-time operators correlation
functions from the knowledge of single-time expectation values, which in
turn implies the knowledge of the density matrix evolution \cite%
{carmichael,cohen,loudon}. The importance of this theorem comes from the
physical information contained in the operator correlations. In fact, in a
stationary regime, it is possible to relate the Fourier transform of these
objects with the spectrum of the decay process \cite{blum}. Furthermore, in
radiant systems, the statistic of the scattered field can be described
through system operator correlations~\cite{carmichael,cohen,loudon}.

Another central cornerstone of non-equilibrium quantum Markovian dynamics is
the quantum detailed balance condition, which imposes severe symmetry
properties on the operator correlations structure. While in classical
stochastic processes this condition has a clear meaning in terms of
transitions between the available states of the system \cite{kampen,klein},
in quantum dissipative systems this condition relies in the time reversal
property of the underlying stationary microscopic Hamiltonian evolution~\cite%
{carmichaelZ,carmichaelPRA,agarwalS,agarwal,alickipaper,verri}. The
breakdown of this condition has direct experimental implications \cite%
{denisov}.

Although the applicability of the Markovian approximation range over many
physical situations \cite{alicki,nielsen,blum,carmichael,cohen,loudon},
there exist several real systems whose dynamics present strong departures
from it. Remarkable examples are anomalous intermittent fluorescence in
quantum dots \cite{michler,schlegel,brokmann,grigolini}, the presence of $%
1/f $ noise in phase and charge superconducting qubits \cite%
{makhlinReport,falci}, and band gap materials \cite{john,yablo}.

Consistently with the existence of experimental situations that can not be
described by a Markovian evolution, in the context of different approaches
recent effort was dedicated to characterize non-Markovian operator
correlation\ dynamics \cite{lang,yan,alonso}.

While the description of non-Markovian processes may depends on each
specific situation, there exists an increasing interest in describing these
kind of processes by introducing memory contributions in standard Lindblad
evolutions \cite%
{barnett,wilkie,budini,cresser,lidar,sabrina,maniscalco,jpa,gbma}. This
procedure provide easy manageable equations. Nevertheless, this technique
does not have associated a rule for calculating operator correlations.

In this paper we explore the possibility of establishing a QRT for
non-Markovian master equations that can be cast in the form of Lindblad
equations with memory contributions \cite%
{barnett,wilkie,budini,cresser,lidar,sabrina,maniscalco,jpa,gbma}. We will
base our considerations in a generalized Born-Markov approximation (GBMA) 
\cite{gbma}, which allows us to develop our results from an underlying
microscopic Hamiltonian description.

The paper is outlined as follows. In Section II we review the derivation of
the GBMA from a full microscopic description. Based on this approach, in
Section III we search the conditions under which a non-Markovian QRT can be
established. In Section IV we relate the non-Markovian QRT with a detailed
balance condition. In Section V we exemplify our theoretical results by
analyzing the correlation dynamics of a two level system embedded in a
complex thermal environment described in a GBMA. In section VI we give the
conclusions.

\section{Generalized Born-Markov approximation}

The GBMA applies for complex structured environments whose action over the
system can be well approximated by a direct sum of sub-reservoirs, each one
being able to induce by itself a Markovian system dynamics. Under this
condition, the system evolution can be written as a Lindblad equation
characterized by a random dissipative rate \cite{gbma}.

Here, we review the microscopic derivation of this approximation by using a
well known projector operator technique \cite{haake,breuer}. This equivalent
derivation is useful for clarifying that the GBMA is not restricted to a
second order approximation. In fact, the projector technique provides a
rigorous procedure that allows to obtain the system dynamics up to any
desired order in the interaction Hamiltonian.

We assume a full microscopic Hamiltonian description of the interaction of a
system $S$ with its environment $B$%
\begin{equation}
H_{T}=H_{S}+H_{B}+H_{I}.  \label{total}
\end{equation}%
Here, $H_{S}$ and $H_{B}$ correspond to the system and bath Hamiltonians
respectively. The term $H_{I}=q_{S}\otimes Q_{B}$\ describes their mutual
interaction, with the operators $q_{S}$ and $Q_{B}$ acting on the system and
bath Hilbert spaces respectively.

The system density matrix follows after eliminating the environment degrees
of freedom, $\rho _{S}(t)=\mathrm{Tr}_{B}\{\rho _{T}(t)\},$ where the total
density matrix $\rho _{T}(t)$\ evolves as%
\begin{equation}
\frac{d\rho _{T}(t)}{dt}=\frac{-i}{\hbar }[H_{T},\rho _{T}(t)]\equiv 
\mathcal{L}_{T}[\rho _{T}(t)].  \label{rho}
\end{equation}%
The GBMA \cite{gbma} can be derived by introducing the projector $\mathcal{P}
$ defined by%
\begin{equation}
\mathcal{P}\rho _{T}(t)\equiv \sum_{R}\rho _{R}(t)\otimes \Xi _{R},
\label{projector}
\end{equation}%
where $\Xi _{R}$ is given by%
\begin{equation}
\Xi _{R}\equiv \Pi _{R}\rho _{B}\Pi _{R},
\end{equation}%
with $\rho _{B}$ being the stationary state of the bath, while the system
states $\rho _{R}(t)$ are defined by%
\begin{equation}
\rho _{R}(t)\equiv \frac{\mathrm{Tr}_{B}\{\Pi _{R}\rho _{T}(t)\Pi _{R}\}}{%
\mathrm{Tr}_{B}\{\Pi _{R}\rho _{B}\Pi _{R}\}}.  \label{rhoR}
\end{equation}%
Here, we have introduced a set of projectors $\Pi _{R}=\sum_{\{\epsilon
_{R}\}}\left\vert \epsilon _{R}\right\rangle \left\langle \epsilon
_{R}\right\vert ,$ which provides an orthogonal decomposition of the unit
operator [$I_{B}$] in the Hilbert space of the bath, $\sum_{R}\Pi _{R}=I_{B},
$ with $\Pi _{R}\Pi _{R^{\prime }}=\Pi _{R}\delta _{R,R^{\prime }}.$ The
full set of states $\left\vert \epsilon _{R}\right\rangle $ corresponds to
the base where $\rho _{B}$\ is diagonal, which implies $\sum_{R}\Xi
_{R}=\rho _{B}.$

It is easy to realize that $\mathcal{P}^{2}=\mathcal{P}.$ In physical terms,
this projector takes in account that each bath-subspace associated to the
projectors $\Pi _{R}$ induces a different system dynamics, each one
represented by the states $\rho _{R}(t).$ On the other hand, notice that the
standard projector $\mathcal{P}\rho _{T}(t)=\mathrm{Tr}_{B}\{\rho
_{T}(t)\}\otimes \rho _{B}=\rho _{S}(t)\otimes \rho _{B}$ \cite{haake,breuer}%
, is recuperated when all the states $\rho _{R}(t)$ have the same dynamics.

From Eq.~(\ref{projector}), the system density matrix follows as%
\begin{equation}
\rho _{S}(t)=\mathrm{Tr}_{B}\{\mathcal{P}\rho _{T}(t)\}=\sum_{R}P_{R}\rho
_{R}(t)\equiv \langle \rho _{R}(t)\rangle .  \label{sum}
\end{equation}%
This equation defines the system state as an average over the density
matrixes $\rho _{R}(t),$ each one participating with weight $P_{R}.$ These
parameters are defined by the weight of each subspace in the full stationary
bath state%
\begin{equation}
P_{R}=\mathrm{Tr}_{B}\{\Xi _{R}\}=\mathrm{Tr}_{B}\{\Pi _{R}\rho
_{B}\}=\sum_{\{\epsilon _{R}\}}\left\langle \epsilon _{R}\right\vert \rho
_{B}\left\vert \epsilon _{R}\right\rangle ,  \label{pesos}
\end{equation}%
which in consequence satisfy $\sum_{R}P_{R}=1.$

By writing the evolution Eq.~(\ref{rho}) in an interaction representation,
and splitting the full dynamics in contributions $\mathcal{P}\rho _{T}(t)$
and $\mathcal{Q}\rho _{T}(t),$ where $\mathcal{Q}=1-\mathcal{P},$ up to
second order in the interaction Hamiltonian it follows \cite{haake,breuer}%
\begin{equation}
\frac{d\mathcal{P}\rho _{T}(t)}{dt}=\int_{0}^{t}dt^{\prime }\mathcal{PL}%
_{T}(t)\mathcal{L}_{T}(t^{\prime })\mathcal{P}\rho _{T}(t^{\prime }),
\label{Interaction}
\end{equation}%
where $\mathcal{L}_{T}(t)$ is the total Liouville operator in a interaction
representation. Here, we have assumed an uncorrelated initial state, $\rho
_{T}(0)=\rho _{S}(0)\otimes \rho _{B}.$

By assuming an interaction Hamiltonian with a direct sum structure%
\begin{equation}
H_{I}=H_{I_{1}}\oplus H_{I_{2}}\cdots \oplus H_{I_{R}}\oplus
H_{I_{R+1}}\cdots ,  \label{suma}
\end{equation}%
where each term satisfies $H_{I_{R}}=\Pi _{R}H_{I}\Pi _{R},$ from Eq.~(\ref%
{Interaction}) it follows that each state $\rho _{R}(t),$ in a Schr\"{o}%
dinger representation, evolves as%
\begin{eqnarray}
\frac{d\rho _{R}(t)}{dt} &=&\frac{-i}{\hbar }[H_{S},\rho _{R}(t)]-\left( 
\frac{1}{\hbar }\right) ^{2}\int_{0}^{\infty }dt^{\prime }  \label{markov} \\
&&\mathrm{Tr}_{B_{R}}\{[H_{I_{R}},[H_{I_{R}}(-t^{\prime }),\rho
_{R}(t)\otimes \rho _{B_{R}}]].  \notag
\end{eqnarray}%
with $\rho _{B_{R}}\equiv \Xi _{R}/P_{R},$ and where $\mathrm{Tr}%
_{B_{R}}\{\bullet \}\equiv \mathrm{Tr}_{B}\{\Pi _{R}\bullet \Pi _{R}\}.$ The
corresponding initial condition reads $\rho _{R}(0)=\rho _{S}(0),$ which
follows from Eq.~(\ref{rhoR}). Furthermore, in this evolution we have
introduced a Markov approximation, which applies when each bath subspace
corresponding to the projectors $\Pi _{R}$ defines a Markovian
sub-environment.

The evolution Eq.~(\ref{markov}), disregarding transients of the order of
the sub-bath correlation time, can be always well approximated by a Lindblad
equation \cite{alicki}%
\begin{equation}
\frac{d\rho _{R}(t)}{dt}=\mathcal{L}_{H}[\rho _{R}(t)]+\gamma _{R}\mathcal{L}%
[\rho _{R}(t)],  \label{LindbladRandom}
\end{equation}%
where $\mathcal{L}_{H}[\bullet ]=-(i/\hbar )[H_{S},\bullet ],$ and the
dissipative contribution is defined by a Lindblad superoperator \cite{alicki}%
\begin{equation}
\mathcal{L}[\bullet ]=\frac{1}{2}\sum_{\alpha \beta }a_{\alpha \beta
}([V_{\alpha },\bullet V_{\beta }^{\dagger }]+[V_{\alpha }\bullet ,V_{\beta
}^{\dagger }]).  \label{lindblad}
\end{equation}%
Here, the set of system operators $\{V_{\alpha }\}$ and the dimensionless
Hermitian matrix $a_{\alpha \beta }$ depend on the underlying microscopic
interaction. The rates $\gamma _{R}$ follow from a Fermi golden rule when
applied to the manifold of states $\{\left\vert \epsilon _{R}\right\rangle
\} $ that define each Markovian sub-reservoir.

While the density matrixes $\rho _{R}(t)$ follow a Markovian evolution, the
system state $\rho _{S}(t)$ evolves with a completely positive \cite%
{alicki,nielsen} non-Markovian evolution, property inherited from the random
Lindblad structure Eq.~(\ref{LindbladRandom}). The average of this equation
over the set $\{\gamma _{R},P_{R}\}$ can be performed in a Laplace domain,
from where it follows%
\begin{equation}
\frac{d\rho _{S}(t)}{dt}=\mathcal{L}_{H}[\rho _{S}(t)]+\int_{0}^{t}d\tau \,%
\mathbb{L}(t-\tau )[\rho _{S}(\tau )],  \label{noMarkovMaster}
\end{equation}%
where the superoperator $\mathbb{L}(t)$ is defined by the relation%
\begin{equation}
\langle G_{R}(u)\gamma _{R}\mathcal{L}\rangle \lbrack \bullet ]=\langle
G_{R}(u)\rangle \mathbb{L}(u)[\bullet ].  \label{memory}
\end{equation}%
Here, $u$ is the Laplace variable and $G_{R}(u)$ is the Markovian propagator
of each state $\rho _{R}(t),$ i.e., $G_{R}(u)\equiv \lbrack u-(\mathcal{L}%
_{H}+\gamma _{R}\mathcal{L)}]^{-1}.$ Depending on the set $\{\gamma
_{R},P_{R}\},$ which specify the complex environment, Eq.~(\ref%
{noMarkovMaster}) may lead to a reach variety of system decay behaviors as
well as to many different structures of \textit{non-local Lindblad equations.%
} In fact, in general $\mathbb{L}(u)$ consists in a sum of Lindblad terms,
each one characterized a different memory kernel.

The structure of the superoperator $\mathbb{L}(u)$ can be simplified in an
effective approximation \cite{gbma}, which consists of discarding the
dependence introduced by the Lindblad superoperator $\mathcal{L}$ in the
propagator $G_{R}(u)$, i.e., $\mathcal{L}_{R}\rightarrow -\mathrm{I}$\textrm{%
.} From Eq.~(\ref{memory}) it follows the approximated solution $\mathbb{L}%
(u)\simeq K(u-\mathcal{L}_{H})\mathcal{L},$ which implies the evolution%
\begin{equation}
\frac{d\rho _{S}(t)}{dt}\simeq \mathcal{L}_{H}[\rho
_{S}(t)]+\int_{0}^{t}d\tau K(t-\tau )e^{(t-\tau )\mathcal{L}_{H}}\mathcal{L}%
[\rho _{S}(\tau )].  \label{aproximada}
\end{equation}%
with $K(u)=\langle \gamma _{R}(u+\gamma _{R})^{-1}\rangle \langle (u+\gamma
_{R})^{-1}\rangle ^{-1}.$ If the time scale of the unitary dynamics is
larger than the time scale of the memory kernel, the unitary contribution
can be discarded leading to a single memory Lindblad equation. We remark
that structures similar to Eq.~(\ref{aproximada}) were obtained in the
context of other approaches \cite%
{wilkie,budini,cresser,lidar,sabrina,maniscalco,jpa}. The GBMA, here defined
through the projector Eq.~(\ref{projector}), allows us to associate an
underlying well defined microscopic description to these kind of equations.

\section{Quantum regression theorem}

For Markovian master equations the QRT \cite{carmichael,cohen,loudon}
provides a direct relation between the evolution of the expectation values
of system observables and their corresponding correlation functions. Here we
will explore the possibility of formulating an equivalent relation when the
system dynamics can be described through the GBMA.

\subsection{Random rate formulation for operators correlations}

Let us introduce a complete set of operators $\{A_{\mu }\}$ of the system,
collected into a vector $\mathbf{A}$, and consider the expectation values%
\begin{equation}
\overline{\mathbf{A}(t)}\equiv \mathrm{Tr}_{SB}[\mathbf{A}(t)\rho _{T}(0)],
\label{average}
\end{equation}%
as well as the correlation functions%
\begin{equation}
\overline{O(t)\mathbf{A}(t+\tau )}\equiv \mathrm{Tr}_{SB}[O(t)\mathbf{A}%
(t+\tau )\rho _{T}(0)],  \label{correlation}
\end{equation}%
where $O(t)$ is an arbitrary system operator. The time dependence of the
operators refers to a Heisenberg representation with respect to the total
Hamiltonian Eq.~(\ref{total}), i.e., $O(t)=\exp [(i/\hbar )tH_{T}]O(0)\exp
[-(i/\hbar )tH_{T}].$

From Eq.~(\ref{sum}), we can write the expectation values as an average over
the solutions corresponding to each rate%
\begin{equation}
\overline{\mathbf{A}(t)}=\left\langle \mathrm{Tr}_{S}[\mathbf{A}(0)\rho
_{R}(t)]\right\rangle \equiv \langle \overline{\mathbf{A}(t)}_{R}\rangle .
\end{equation}%
In order to work out the operator correlations, we first express the total
initial density matrix as $\rho _{T}(0)~=~\exp [(i/\hbar )tH_{T}]\rho
_{T}(t)\exp [-(i/\hbar )tH_{T}]$. Then, by using the cyclic property of the
trace, from Eq.~(\ref{correlation}) we obtain 
\begin{equation}
\overline{O(t)\mathbf{A}(t+\tau )}=\mathrm{Tr}_{S}\{\mathbf{A}(0)\mathrm{Tr}%
_{B}[O_{SB}(\tau )]\},  \label{correlation1}
\end{equation}%
where the operator $O_{SB}(\tau )$ satisfies 
\begin{equation}
\frac{d}{d\tau }O_{SB}(\tau )=-\frac{i}{\hbar }[H_{T},O_{SB}(\tau )],
\end{equation}%
with $\left. O_{SB}(\tau )\right\vert _{\tau =0}=\rho _{T}(t)O(0).$ This
system-bath operator evolves as the total density matrix, Eq.~(\ref{rho}).
On the other hand, Eq.~(\ref{projector}) allows us to write the initial
condition as $\left. O_{SB}(\tau )\right\vert _{\tau =0}\approx
\sum_{R}[\rho _{R}(t)O(0)]\otimes \Xi _{R}$ \cite{footnoteCero}. Therefore,
the reduced dynamics of $O_{SB}(\tau )$ can also be described in a GBMA,
which deliver%
\begin{equation}
\mathrm{Tr}_{B}[O_{SB}(\tau )]=\langle \exp [(\mathcal{L}_{H}+\mathcal{L}%
_{R})\tau ]\rho _{R}(t)\rangle O(0),
\end{equation}%
where, for shortening the notation we defined $\mathcal{L}_{R}\equiv \gamma
_{R}\mathcal{L}$. From Eq.~(\ref{correlation1}), it follows%
\begin{eqnarray}
\overline{O(t)\mathbf{A}(t+\tau )} &=&\langle \mathrm{Tr}_{S}\{\mathbf{A}%
(0)e^{(\mathcal{L}_{H}+\mathcal{L}_{R})\tau }[\rho _{R}(t)O(0)]\}\rangle 
\notag \\
&\equiv &\langle \overline{O(t)\mathbf{A}(t+\tau )}_{R}\rangle .
\label{QRT1}
\end{eqnarray}%
This expression is an average over the random set $\{\gamma _{R},P_{R}\}$ of
the corresponding Markovian correlation expressions \cite{carmichael}. This
characteristic provides us a central result, which allows us to extend the
averaging procedure [Eq.~(\ref{sum})] corresponding to the GBMA for operator
correlations as well. In fact, \textit{higher correlations operators can
also be obtained as an average, over the random rate set, of the Markovian
expressions corresponding to each state }$\rho _{R}(t).$ For example, using
the same steps as before, for arbitrary system operators $O_{1}$ and $O_{2}$%
, it is possible to obtain%
\begin{eqnarray}
\overline{O_{1}(t)\mathbf{A}(t+\tau )O_{2}(t)} &=&\langle \mathrm{Tr}_{S}\{%
\mathbf{A}e^{(\mathcal{L}_{H}+\mathcal{L}_{R})\tau }[O_{2}\rho
_{R}(t)O_{1}]\}\rangle  \notag \\
&\equiv &\langle \overline{O_{1}(t)\mathbf{A}(t+\tau )O_{2}(t)}_{R}\rangle ,
\label{QRT2}
\end{eqnarray}%
which also correspond to an average over the corresponding Markovian
dynamics \cite{carmichael}.

\subsection{Expectation and correlation evolution}

From the previous result, we can write the evolution of both, expectation
values and correlations, as an average over the random rate set 
\begin{subequations}
\begin{eqnarray}
\frac{d}{dt}\overline{\mathbf{A}(t)} &=&\langle \mathbf{\hat{M}}_{R}%
\overline{\mathbf{A}(t)}_{R}\rangle ,  \label{onepoint} \\
\frac{d}{d\tau }\overline{O(t)\mathbf{A}(t+\tau )} &=&\langle \mathbf{\hat{M}%
}_{R}\overline{O(t)\mathbf{A}(t+\tau )}_{R}\rangle .  \label{twopoint}
\end{eqnarray}
Here, the matrix $\mathbf{\hat{M}}_{R}$ acts on the indices of $\mathbf{A}$
and is defined by the condition 
\end{subequations}
\begin{equation}
\mathrm{Tr}_{S}\{\mathbf{A}(\mathcal{L}_{H}+\mathcal{L}_{R})[O]\}=\mathbf{%
\hat{M}}_{R}\mathrm{Tr}_{S}\{\mathbf{A}O\}.
\end{equation}%
When $\gamma _{R}$ is fixed, the evolution equations (\ref{onepoint}) for
expectation values and (\ref{twopoint}) for correlation functions are
identical, which recovers the \textit{QRT for Markovian dynamics}. In the
non-Markovian case, however, both equations still involve the average over
the dissipation rate.

As for the density matrix \cite{gbma}, the averaged evolutions can be worked
out in the Laplace domain. The expectation value can be expressed as $%
\overline{\mathbf{A}(u)}=\langle \mathbf{\hat{G}}_{R}(u)\overline{\mathbf{A}%
(0)}\rangle $, with the matrix propagator $\mathbf{\hat{G}}_{R}(u)\equiv (u+%
\mathbf{\hat{M}}_{R})^{-1}.$ After introducing the identity operator in the
form $\overline{\mathbf{A}(u)}=\langle \mathbf{\hat{G}}_{R}(u)(u+\mathbf{%
\hat{M}}_{R})\rangle ^{-1}\langle \mathbf{\hat{G}}_{R}(u)\overline{\mathbf{A}%
(0)}\rangle $, we arrive to the deterministic closed evolution 
\begin{subequations}
\begin{equation}
\frac{d}{dt}\overline{\mathbf{A}(t)}=-\int_{0}^{t}dt^{\prime }\mathbb{\hat{M}%
}(t-t^{\prime })\overline{\mathbf{A}(t^{\prime })}.  \label{promedio}
\end{equation}%
Using a similar procedure, for the correlation we get 
\begin{eqnarray}
\frac{d}{d\tau }\overline{O(t)\mathbf{A}(t+\tau )} &=&-\int_{0}^{\tau
}dt^{\prime }\ \mathbb{\hat{M}}(\tau -t^{\prime })\overline{O(t)\mathbf{A}%
(t+t^{\prime })}  \notag \\
&&+\mathbf{I}(t,\tau ).  \label{correlacion}
\end{eqnarray}
The deterministic kernel matrix $\mathbb{\hat{M}}(t)$ fulfills the equation 
\end{subequations}
\begin{equation}
\mathbb{\hat{M}}(u\mathbb{)=}\langle \mathbf{\hat{G}}_{R}(u)\rangle
^{-1}\langle \mathbf{\hat{G}}_{R}(u)\mathbf{\hat{M}}_{R}\rangle ,
\label{kernel}
\end{equation}%
while the inhomogeneous term $\mathbf{I}(t,\tau )$ is defined by%
\begin{equation}
\mathbf{I}(t,u)\mathbb{=}\langle \mathbf{\hat{G}}_{R}(u)\rangle ^{-1}\langle 
\mathbf{\hat{G}}_{R}(u)\overline{O(t)\mathbf{A}(t)}_{R}\rangle -\langle 
\overline{O(t)\mathbf{A}(t)}_{R}\rangle .
\end{equation}%
Besides that Eq.~(\ref{twopoint}) has the same structure as Eq.~(\ref%
{onepoint}), the inhomogeneous term is only present in the correlation
evolution, Eq.~(\ref{correlacion}). $\mathbf{I}(t,\tau )$ arise because the
initial condition of each contribution in Eq.~(\ref{twopoint}) is correlated
with respect to its propagator. In fact, notice that both $\mathbf{\hat{G}}%
_{R}(u)$ and $\overline{O(t)\mathbf{A}(t)}_{R}$ depend on $\gamma _{R}$,
which implies that theses objects are correlated with respect to the random
rate statistics. The dependence of $\overline{O(t)\mathbf{A}(t)}_{R}$ on $%
\gamma _{R}$ follows from $\overline{O(t)\mathbf{A}(t)}_{R}=\mathrm{Tr}%
_{S}\{O(0)\mathbf{A}(0)\rho _{R}(t)].$ On the other hand, as Eq.~(\ref%
{onepoint}) is defined with initial conditions fixed at $t=0$, its initial
condition $\overline{\mathbf{A}(0)}_{R}$\ does not depends on $\gamma _{R},$
which in turn implies that the inhomogeneous term is not present in the
averaged evolution Eq.~(\ref{promedio}).

Due to the inhomogeneous term $\mathbf{I}(t,\tau ),$ the QRT is not
fulfilled in general. A non-Markovian QRT is only valid when this term
vanish, which leads to the condition%
\begin{equation}
\langle \mathbf{\hat{G}}_{R}(u)\overline{O(t)\mathbf{A}(t)}_{R}\rangle 
\overset{QRT}{=}\langle \mathbf{\hat{G}}_{R}(u)\rangle \langle \overline{O(t)%
\mathbf{A}(t)}_{R}\rangle  \label{condition}
\end{equation}%
This equality is always satisfied for Markovian dynamics because the average
over the dissipation rate is absent. We also note that \textit{a
non-Markovian QRT can be asymptotically valid if the stationary state }$\rho
_{R}^{\infty }\equiv \rho _{R}(\infty )$\textit{\ does not depend on }$%
\gamma _{R}$ \cite{artificial}. In fact, in this situation $%
\lim_{t\rightarrow \infty }\overline{O(t)\mathbf{A}(t)}_{R}=\mathrm{Tr}%
_{S}\{O(0)\mathbf{A}(0)\rho _{R}^{\infty }]$ is independent of $\gamma _{R},$
and then the condition Eq.~(\ref{condition}) is automatically satisfied.
However, if the asymptotic state $\rho _{R}^{\infty }$ depends on $\gamma
_{R}$ the inhomogeneous term will contribute at all times, even in the
asymptotic regime, and the QRT is invalidated. The same condition is valid
for higher operators correlations.

\subsection{Non-Markovian dynamics}

The evolution Eq.~(\ref{promedio}) and (\ref{correlacion}) can be formally
integrated in the Laplace domain. For the expectation values we get 
\begin{subequations}
\label{noMarkov}
\begin{equation}
\overline{\mathbf{A}(t)}=\mathbb{\hat{G}}(t)\overline{\mathbf{A}(0)},
\label{solution1}
\end{equation}
while for the correlations it follows 
\begin{equation}
\overline{O(t)\mathbf{A}(t+\tau )}=\mathbb{\hat{G}}(\tau )\overline{O(t)%
\mathbf{A}(t)}+\mathbf{F}(t,\tau ).  \label{solution}
\end{equation}
The non-Markovian propagator is defined by 
\end{subequations}
\begin{equation}
\mathbb{\hat{G}}(u)=\frac{1}{u+\mathbb{\hat{M}}(u\mathbb{)}},
\end{equation}%
and the extra inhomogeneous term is%
\begin{equation}
\mathbf{F}(t,\tau )=\langle \mathbf{\hat{G}}_{R}(\tau )\overline{O(t)\mathbf{%
A}(t)}_{R}\rangle -\langle \mathbf{\hat{G}}_{R}(\tau )\rangle \langle 
\overline{O(t)\mathbf{A}(t)}_{R}\rangle .  \label{extra}
\end{equation}%
These expressions explicitly show that the departure from condition Eq.~(\ref%
{condition}) measures the size of the dynamical effects which can not be
captured by assuming valid the QRT. In fact, the QRT is fulfilled only when $%
\mathbf{F}(t,\tau )$ vanishes.

Eq.~(\ref{solution1}) and (\ref{solution}) are consistent with the averaging
procedure over Markovian solutions. In fact, they can be expressed as $%
\overline{\mathbf{A}(t)}=\langle \mathbf{\hat{G}}_{R}(\tau )\rangle 
\overline{\mathbf{A}(0)},$ and for the correlations as%
\begin{equation}
\overline{O(t)\mathbf{A}(t+\tau )}=\langle \mathbf{\hat{G}}_{R}(\tau )%
\overline{O(t)\mathbf{A}(t)}_{R}\rangle ,  \label{propagada}
\end{equation}%
which\ in fact are an average over Markovian solutions.

\subsection{Fluctuation operators}

Of special interest is to study the correlation dynamics of fluctuation
operators, which are defined as the departure from expectation values 
\begin{subequations}
\begin{eqnarray}
\delta \!\mathbf{A}(t) &\equiv &\mathbf{A}(t)-\overline{\mathbf{A}(t)}, \\
\delta \!O(t) &\equiv &O(t)-\overline{O(t)}.
\end{eqnarray}
For Markovian dynamics the correlation of these operators also satisfies a
QRT. These objects are relevant to split the spectrum, defined as the
Fourier transform of the stationary correlations, in a coherent and
incoherent components~\cite{carmichael}.

From Eq.~(\ref{propagada}) we can write 
\end{subequations}
\begin{eqnarray}
\overline{O(t)\mathbf{A}(t+\tau )} &=&\langle \mathbf{\hat{G}}_{R}(\tau )%
\overline{\delta \!O(t)\delta \!\mathbf{A}(t+\tau )}_{R}\rangle  \notag \\
&&+\langle \overline{O(t)}_{R}\overline{\mathbf{A}(t+\tau )}_{R}\rangle .
\label{fluctuation}
\end{eqnarray}%
For Markovian evolutions, in the asymptotic time regime $(t\rightarrow
\infty )$, the first contribution can be associated with the incoherent
spectrum component while the second one, after taking the extra limit $\tau
\rightarrow \infty $, with the coherent spectrum part. After averaging over
the random rate, these associations remains valid for the non-Markovian
case. In particular, we note that the coherent component%
\begin{equation}
\lim_{\substack{ t\rightarrow \infty  \\ \tau \rightarrow \infty }}\overline{%
O(t)\mathbf{A}(t+\tau )}=\langle \overline{O(\infty )}_{R}\ \overline{%
\mathbf{A}(\infty )}_{R}\rangle  \label{coherente}
\end{equation}%
is an average of the corresponding Markovian contributions.

The correlation Eq.~(\ref{propagada}) can also be written as%
\begin{equation}
\overline{O(t)\mathbf{A}(t+\tau )}=\overline{\delta \!O(t)\delta \!\mathbf{A}%
(t+\tau )}+\overline{O(t)}\ \ \overline{\mathbf{A}(t+\tau )}.  \label{vale1}
\end{equation}%
As for the Markovian case, this expression follows immediately from the
microscopic definition Eq.~(\ref{correlation}). From this relation and Eq.~(%
\ref{fluctuation}), we get%
\begin{equation}
\overline{\delta \!O(t)\delta \!\mathbf{A}(t+\tau )}=\mathbb{\hat{G}}(\tau )%
\overline{\delta \!O(t)\delta \!\mathbf{A}(t)}+\delta \!\mathbf{F}(t,\tau ).
\label{vale2}
\end{equation}%
Here, the first contribution follows from the QRT when assumed valid for
fluctuations operators, and the second one measures the departure from it,
being defined by%
\begin{eqnarray*}
\delta \!\mathbf{F}(t,\tau )\! &=&\!\langle \mathbf{\hat{G}}_{R}(\tau )%
\overline{\delta \!O(t)\delta \!\mathbf{A}(t)}_{R}\rangle \!-\!\langle 
\mathbf{\hat{G}}_{R}(\tau )\rangle \!\langle \overline{\delta \!O(t)\delta \!%
\mathbf{A}(t)}_{R}\rangle \  \\
&&\!+\langle \overline{O(t)}_{R}\overline{\mathbf{A}(t+\tau )}_{R}\rangle
-\langle \overline{O(t)}_{R}\rangle \langle \overline{\mathbf{A}(t+\tau )}%
_{R}\rangle .
\end{eqnarray*}%
As for operators, in the asymptotic regime the QRT is also valid for
fluctuation operators if the stationary sate $\rho _{R}^{\infty }$ does not
depend on the random rate, which is fact implies $\delta \!\mathbf{F}(\infty
,\tau )=0.$

From Eq.~(\ref{vale1}) and Eq.~(\ref{vale2}), we notice that by assuming
valid the QRT, the coherent spectrum component reads%
\begin{equation}
\lim_{\substack{ t\rightarrow \infty  \\ \tau \rightarrow \infty }}\overline{%
O(t)\mathbf{A}(t+\tau )}\overset{QRT}{=}\langle \overline{O(\infty )}%
_{R}\rangle \langle \overline{\mathbf{A}(\infty )}_{R}\rangle .
\label{coherenteQRT}
\end{equation}%
This expression and Eq.~(\ref{coherente}) indicate that the predictions of
the QRT will differ from the exact dynamics not only in the transient
dynamical behaviors but in general also in the asymptotic correlation
values. In the next sections, we will use the difference between these two
expressions as a measure of the deviation from the validity of the QRT in
the stationary regime.

\section{Detailed balance condition}

In the context of the GBMA, in the previous section we have demonstrated
that the QRT can be assumed valid in an asymptotic regime if the stationary
state $\rho _{R}^{\infty }$\ corresponding to each Markovian contribution
does not depends on the random rate. Here, we will find an equivalent
condition which does not depends on the approximations used to arrive to the
non-Markovian Lindblad equation. We will demonstrate that the previous
result can be associated with a quantum detailed balance condition \cite%
{agarwalS,agarwal,carmichaelZ,carmichaelPRA,alickipaper,verri}, which in
turn is related with the microrreversibility of the underlying microscopic
dynamics~\cite{carmichaelZ}.

\subsection{Classical conditions}

The concept of detailed balance is well established for classical population
master equations \cite{kampen}%
\begin{equation}
\frac{dp_{n}(t)}{dt}=\gamma _{cl}\Big\{\sum_{m}g_{nm}p_{m}(t)-%
\sum_{m}g_{mn}p_{n}(t)\Big\},
\end{equation}%
where $\gamma _{cl}g_{nm}$ define the hopping rates. The classical detailed
balance condition reads%
\begin{equation}
\gamma _{cl}g_{nm}p_{m}(\infty )=\gamma _{cl}g_{mn}p_{n}(\infty ),
\label{clasica}
\end{equation}%
which has an immediate interpretation in terms of the available stationary
transitions. We note that these relations does not depend on the global rate 
$\gamma _{cl}$. Thus, they impose strong relations between the dimensionless
hopping coefficients $\{g_{nm}\}$ and the stationary populations $%
\{p_{n}(\infty )\}.$ In particular, it is possible to prove that when the
stationary state depends on an arbitrary continuous parameter $\varepsilon ,$
$\{p_{m}(\infty ,\varepsilon )\}$, the hopping coefficients must also to
depend on that parameter,$\ \{g_{mn}(\varepsilon )\}.$ If this is not the
case, the detailed balance condition is violated \cite{footnote1}. This
result can be extended to quantum master equations, $\varepsilon $ being the
random rate, establishing a strong relation between the validity of the QRT
for non-Markovian dynamics and the detailed balance condition.

\subsection{Quantum Markovian conditions}

The detailed balance condition can be generalized for quantum dynamics from
the time reversal property~\cite{carmichaelZ} of the stationary system-bath
dynamics. For an open \textit{Markovian} system, it can be written as an
statement of time symmetry for \textit{stationary} two-time operator
correlations \cite{agarwalS,agarwal,carmichaelZ,carmichaelPRA}%
\begin{equation}
\lim_{t\rightarrow \infty }\overline{O(t+\tau )\mathbf{A}(t)}%
_{R}=\lim_{t\rightarrow \infty }\overline{\widetilde{\mathbf{A}}(t+\tau )%
\widetilde{O}(t)}_{R},  \label{reversal}
\end{equation}%
where $\widetilde{O}(t)$ and $\widetilde{\mathbf{A}}(t)$ represent
time-reversed operators \cite{messiah}. From this equation \cite{agarwal},
it is possible to write an equivalent formulation in the Laplace domain as 
\cite{footnote2}%
\begin{equation}
\rho _{R}^{\infty }\frac{1}{u-(\mathcal{L}_{H}^{\#}+\mathcal{L}_{R}^{\#})}%
[\bullet ]=\frac{1}{u-(\widetilde{\mathcal{L}}_{H}+\widetilde{\mathcal{L}}%
_{R})}[\widetilde{\rho }_{R}^{\infty }\bullet ].  \label{detailed}
\end{equation}%
We have introduced dual and time reversed superoperators~\cite%
{agarwalS,agarwal,carmichaelZ,carmichaelPRA}, which respectively are defined
by $\mathrm{Tr}_{S}\{O\mathcal{L}[\rho ]\}=\mathrm{Tr}_{S}\{\rho \mathcal{L}%
^{\#}[O]\},$ and by $\widetilde{\mathcal{L}[O]}=\widetilde{\mathcal{L}}[%
\widetilde{O}]$ \cite{footnote3}. Equation~(\ref{detailed}) is equivalent to
the conditions 
\begin{subequations}
\label{condiciones}
\begin{eqnarray}
\widetilde{\rho }_{R}^{\infty } &=&\rho _{R}^{\infty }, \\
H_{S}\rho _{R}^{\infty } &=&\rho _{R}^{\infty }H_{S}, \\
\gamma _{R}\rho _{R}^{\infty }\mathcal{L}^{\#}[\bullet ] &=&\gamma _{R}%
\widetilde{\mathcal{L}}[\rho _{R}^{\infty }\bullet ].
\end{eqnarray}
From the second equation and the stationary condition, $\{\mathcal{L}%
_{H}+\gamma _{R}\mathcal{L}\}[\rho _{R}^{\infty }]=0,$ it follows $\mathcal{L%
}[\rho _{R}^{\infty }]=0.$ This condition cannot be satisfied consistently
if the stationary state $\rho _{R}^{\infty }$ depends on $\gamma _{R}$. In
fact, the superoperator $\mathcal{L}$ does not has a continuous parametrized
degenerate null eigen-operator. An equivalent conclusion can be obtained
from the third relation. Then, we deduce that whenever $\rho _{R}^{\infty }$
depends on the random rate $\gamma _{R}$ the detailed balance condition is
violated. Therefore, we can affirm that \textit{if the underlying Markovian
evolution of }$\rho _{R}(t)$\textit{\ satisfies the quantum detailed balance
conditions, Eq.~(\ref{condiciones}), the non-Markovian QRT is valid in the
asymptotic regime.} Equivalently, this statement indicates that when the
non-Markovian QRT is not fulfilled, the detailed balance conditions Eq.~(\ref%
{condiciones}) are also not satisfied.

\subsection{Quantum non-Markovian conditions}

The microrreversibility condition Eq.~(\ref{reversal}) can be trivially
extended to the non-Markovian dynamics as 
\end{subequations}
\begin{equation}
\lim_{t\rightarrow \infty }\overline{O(t+\tau )\mathbf{A}(t)}%
=\lim_{t\rightarrow \infty }\overline{\widetilde{\mathbf{A}}(t+\tau )%
\widetilde{O}(t)}.  \label{microrreversibility}
\end{equation}%
After applying the averaging procedure, from Eqs.~(\ref{reversal}) and (\ref%
{detailed}), we get the equivalent condition%
\begin{equation}
\left\langle \rho _{R}^{\infty }\frac{1}{u-(\mathcal{L}_{H}^{\#}+\mathcal{L}%
_{R}^{\#})}\right\rangle [\bullet ]=\left\langle \frac{1}{u-(\widetilde{%
\mathcal{L}}_{H}+\widetilde{\mathcal{L}}_{R})}[\widetilde{\rho }_{R}^{\infty
}\right\rangle \bullet ].  \label{balance}
\end{equation}%
When the stationary state does not depends on the random rate, $\rho
_{R}^{\infty }=\rho _{S}^{\infty }$, Eq.~(\ref{balance}) leads to the
conditions 
\begin{subequations}
\label{classic}
\begin{eqnarray}
\widetilde{\rho }_{S}^{\infty } &=&\rho _{S}^{\infty }, \\
\rho _{S}^{\infty }\{\mathcal{L}_{H}^{\#}+\mathbb{L}^{\#}(u)\}[\bullet ]
&=&\{\widetilde{\mathcal{L}}_{H}+\widetilde{\mathbb{L}}(u)\}[\rho
_{S}^{\infty }\bullet ],  \label{classic2}
\end{eqnarray}
which must to be valid for any value of the Laplace variable $u$ \cite%
{footnote3bis}. We notice that a similar structure also arises when
formulating the detailed balance condition for non-Markovian classical
Fokker-Planck equations~\cite{agarwal}.

In contrast to the previous conditions [Eq.~(\ref{condiciones})], Eq.~(\ref%
{classic}) do not depends on the approximations or formalism used to derive
the non-Markovian system dynamics. In fact, it only depends on the
superoperator $\mathbb{L}(u)$ that defines the density matrix evolution,
Eq.~(\ref{noMarkovMaster}). In this way we establish a general relation
between the non-Markovian QRT and the non-Markovian quantum detailed balance
condition. We can affirm that, \textit{whenever the non-Markovian quantum
detailed balance conditions Eq.~(\ref{classic}) are satisfied, the
non-Markovian QRT is fulfilled in the asymptotic regime. }The superoperators 
$\mathbb{L}^{\#}(u)$ and $\widetilde{\mathbb{L}}(u)$\ follow from $\mathbb{L}%
(u)$ after replacing all involved superoperators by their dual and time
reversed expressions, respectively. In the context of the GBMA, they\
satisfy Eq.~(\ref{memory}) after replacing $\mathcal{L}_{H}$ and $\mathcal{L}
$ by their dual and time reversed expressions.

As we will exemplify in the next section, a typical situation where the
non-Markovian QRT is broken, even in the stationary regime, is in systems at
thermal equilibrium subject to an external perturbation \cite{klein}. In
fact, it is possible to prove\ that conditions Eqs.~(\ref{classic}) are not
satisfied when a dissipative dynamics that by itself fulfill the detailed
balance condition is subject to the action of an external Hamiltonian field
that does not commutate with the system Hamiltonian. Equivalently, in the
context of the GBMA, the presence of the external perturbation implies that $%
\rho _{R}^{\infty }$ depends on the random rate, which broke the fulfillment
of the Markovian conditions Eq.~(\ref{condiciones}).

\section{Decay in a structured thermal reservoir}

Here we will exemplify our theoretical results by studying a two level
system embedded in a complex structured thermal reservoir whose action can
be described through the GBMA. The system Hamiltonian is 
\end{subequations}
\begin{equation}
H_{S}=\frac{\hbar \omega _{A}}{2}\sigma _{z}+H_{ext}(t),
\end{equation}%
where $\hbar \omega _{A}$ is the difference of energy between the two
levels, denoted by $\left\vert \pm \right\rangle $, and $\sigma _{z}$ is the
z-Pauli matrix. $H_{ext}(t)$ represent an external time dependent field.

The dissipative system dynamics can be defined through the evolution of the
states $\rho _{R}(t)$, which reads%
\begin{equation}
\frac{d\rho _{R}(t)}{dt}=\mathcal{L}_{H}[\rho _{R}(t)]+\gamma _{R}^{\prime }%
\mathcal{L}_{th}[\rho _{R}(t)]+\frac{\gamma _{\Phi }}{2}\mathcal{L}_{\Phi
}[\rho _{R}(t)]\}.  \label{randommatrix}
\end{equation}%
with $\mathcal{L}_{H}[\bullet ]=-(i/\hbar )[H_{S},\bullet ]$. The influence
of the structured thermal reservoir is introduced by the Lindblad
superoperator%
\begin{eqnarray}
\mathcal{L}_{th}[\bullet ] &=&\frac{1+n_{th}}{2}([\sigma ,\bullet \sigma
^{\dagger }]+[\sigma \bullet ,\sigma ^{\dagger }])  \notag \\
&&+\frac{n_{th}}{2}([\sigma ^{\dagger },\bullet \sigma ]+[\sigma ^{\dagger
}\bullet ,\sigma ]),
\end{eqnarray}%
and an arbitrary set ${\{\gamma }^{\prime }{_{R},P_{R}\}}$ of random rates
and weights. $\sigma ^{\dagger }$ and $\sigma $ are the raising and lowering
operators acting on the states $\left\vert \pm \right\rangle $. The
dimensionless constant $n_{th}$ defines the temperature $T$ of the
environment as $\exp [-\hbar \omega _{A}/kT]=n_{th}/(n_{th}+1),$ where $k$
is the Boltzmann constant. We have also considered the action of an extra
dispersive environment which is introduced by the Lindblad superoperator%
\begin{equation}
\mathcal{L}_{\Phi }[\bullet ]=([\sigma _{z},\bullet \sigma _{z}]+[\sigma
_{z}\bullet ,\sigma _{z}])/2,
\end{equation}%
and the single non-random rate $\gamma _{\Phi }$.

\subsection{Free decay dynamics}

First we analyze the case without the external excitation, i.e., $%
H_{ext}(t)=0.$

\textit{Density matrix evolution}: The evolution of the system density
matrix follows from Eq.~(\ref{noMarkovMaster}) and (\ref{memory}). By
denoting the matrix elements as%
\begin{equation}
\rho _{S}(t)=\left( 
\begin{array}{cc}
\Pi _{+}(t) & \Phi _{+}(t) \\ 
\Phi _{-}(t) & \Pi _{-}(t)%
\end{array}%
\right) ,
\end{equation}%
in an interaction representation with respect to $\hbar \omega _{A}\sigma
_{z}/2,$ for the populations we get the evolution%
\begin{equation}
\frac{d}{dt}\Pi _{\pm }(t)=\int_{0}^{t}d\tau K(t-\tau )\{\mp \Pi
_{-}^{eq}\Pi _{+}(\tau )\pm \Pi _{+}^{eq}\Pi _{-}(\tau )\},  \label{popular}
\end{equation}%
while for the coherences we obtain%
\begin{equation}
\frac{d}{dt}\Phi _{\pm }(t)=-\int_{0}^{t}d\tau K_{\Phi }(t-\tau )\Phi _{\pm
}(\tau ).  \label{coherencia}
\end{equation}%
The memory kernel functions are defined by%
\begin{eqnarray}
K(u) &=&\left\langle \frac{\gamma _{R}}{u+\gamma _{R}}\right\rangle
\left\langle \frac{1}{u+\gamma _{R}}\right\rangle ^{-1},
\label{kernelefectivo} \\
K_{\Phi }(u) &=&\left\langle \frac{\gamma _{R}^{\Phi }}{u+\gamma _{R}^{\Phi }%
}\right\rangle \left\langle \frac{1}{u+\gamma _{R}^{\Phi }}\right\rangle
^{-1}.
\end{eqnarray}%
For shortening the notation, we introduced the rates $\gamma _{R}\equiv
\gamma _{R}^{\prime }(1+2n_{th})$ and $\gamma _{R}^{\Phi }\equiv \gamma
_{R}/2+\gamma _{\Phi }$. Furthermore, the dimensionless parameters $\Pi
_{+}^{eq}$ and $\Pi _{-}^{eq}$ are defined by $\Pi _{+}^{eq}/\Pi
_{-}^{eq}=n_{th}/(n_{th}+1)$ and $\Pi _{+}^{eq}+\Pi _{-}^{eq}=1$.

\textit{Quantum detailed balance condition}: In order to check condition
Eq.~(\ref{classic}), we note that the evolutions Eq.~(\ref{popular}) and (%
\ref{coherencia}) can be cast in the superoperator form%
\begin{equation}
\mathbb{L}(u)[\bullet ]=\frac{1}{1+2n_{th}}K(u)\mathcal{L}_{th}[\bullet ]+%
\frac{K_{\phi }(u)}{2}\mathcal{L}_{\Phi }[\bullet ],  \label{lindbladU}
\end{equation}%
where$\ K_{\phi }(u)=K_{\Phi }(u)-K(u)/2.$ The corresponding stationary
state reads%
\begin{equation}
\rho _{S}^{\infty }=\Pi _{+}^{eq}\left\vert +\right\rangle \left\langle
+\right\vert +\Pi _{-}^{eq}\left\vert -\right\rangle \left\langle
-\right\vert ,
\end{equation}%
which due to the time reversal invariance of Hamiltonian eigenvectors
satisfies $\widetilde{\rho }_{S}^{\infty }=\rho _{S}^{\infty }$. Then, it is
easy to prove that Eq.~(\ref{classic}) is satisfied identically.
Consistently, notice that the underlying Markovian dynamic Eq.~(\ref%
{randommatrix}) satisfies the conditions Eq.~(\ref{condiciones}).

\textit{Quantum regression theorem}: As the quantum detailed balance
condition is satisfied, the QRT is valid in an asymptotic regime.
Consistently the stationary state of Eq.~(\ref{randommatrix}) does not
depend on $\gamma _{R}$ $[\rho _{R}^{\infty }=\rho _{S}^{\infty }].$

The transient deviation from the QRT can be easily obtained for this
example. First, we note that the density matrix evolution defined by Eq.~(%
\ref{lindbladU}) is equivalent to the non-Markovian Bloch equation 
\begin{subequations}
\begin{eqnarray}
\frac{dS_{X}(t)}{dt} &=&-\int_{0}^{t}d\tau K_{\Phi }(t-\tau )S_{X}(\tau ), \\
\frac{dS_{Y}(t)}{dt} &=&-\int_{0}^{t}d\tau K_{\Phi }(t-\tau )S_{Y}(\tau ), \\
\frac{dS_{Z}(t)}{dt} &=&-\int_{0}^{t}d\tau K(t-\tau )[S_{Z}(\tau )-\mathcal{S%
}_{Z}^{\infty }],
\end{eqnarray}
where $S_{j}(t)\equiv \mathrm{Tr}_{S}\{\rho _{S}(t)\sigma _{j}\}$ are the
expectation values of the Pauli matrixes $\sigma _{j},$ and $\mathcal{S}%
_{Z}^{\infty }\equiv \Pi _{+}^{eq}-\Pi _{-}^{eq}.$ In order to deal with
diagonal matrixes, we analyze the correlations in the base $\mathbf{A}%
=\{\sigma _{x},\sigma _{y},(\sigma _{z}-\mathcal{S}_{Z}^{\infty }),\mathrm{I}%
\}$. Then, the propagator for operator expectation values, $\overline{%
\mathbf{A}(t)}=\mathbb{\hat{G}}(t)\overline{\mathbf{A}(0)},$ can be written
as 
\end{subequations}
\begin{equation}
\mathbb{\hat{G}}(t)=\mathrm{diag}\{P_{\Phi }(t),P_{\Phi }(t),P_{\Pi }(t),1\}.
\label{diagonal}
\end{equation}%
Here, we defined the functions $P_{\Pi }(u)=[u+K(u)]^{-1}$ and $P_{\Phi
}(u)=[u+K_{\Phi }(u)]^{-1},$ which in term of the random rate set can be
written in the time domain as%
\begin{equation}
P_{\Pi }(t)=\left\langle \exp [-\gamma _{R}t]\right\rangle ,\ \ \ \ \ \ \ \
P_{\Phi }(t)=e^{-\gamma _{\Phi }t}P_{\Pi }(t/2).  \label{survivals}
\end{equation}%
On the other hand, the extra inhomogeneous term [Eq.~(\ref{extra})] that
defines the operator correlations, $\overline{O(t)\mathbf{A}(t+\tau )}=%
\mathbb{\hat{G}}(\tau )\overline{O(t)\mathbf{A}(t)}+\mathbf{F}(t,\tau ),$
can be written~as%
\begin{equation}
\mathbf{F}(t,\tau )=\mathbb{\hat{G}}_{\Pi }(t,\tau )\mathbf{F}_{\Pi }+%
\mathbb{\hat{G}}_{\Phi }(t,\tau )\mathbf{F}_{\Phi },  \label{Fextra}
\end{equation}%
where we have defined the vectors 
\begin{subequations}
\begin{eqnarray}
\mathbf{F}_{\Pi } &=&\mathrm{Tr}_{S}[O(0)\mathbf{A}(0)\{\rho
_{S}^{+}(0)-\rho _{S}^{\infty }\}], \\
\mathbf{F}_{\Phi } &=&\mathrm{Tr}_{S}[O(0)\mathbf{A}(0)\rho _{S}^{-}(0)],
\end{eqnarray}
with $\rho _{S}^{\pm }(0)\equiv \lbrack \rho _{S}(0)\pm \sigma _{z}\rho
_{S}(0)\sigma _{z}]/2$ \cite{footnote4}. We note that $\mathbf{F}_{\Pi }$
measure the departure of the initial populations from the equilibrium values 
$\Pi _{\pm }^{eq},$ while $\mathbf{F}_{\Phi }$ measure the departure of the
initial coherences from their null stationary value. Thus, $\mathbf{F}%
(t,\tau )$ vanishes if the system start in the equilibrium state $\rho
_{S}^{\infty }$. On the other hand, the time dependence of $\mathbf{F}%
(t,\tau )$ is defined by the matrixes 
\end{subequations}
\begin{subequations}
\begin{eqnarray}
\mathbb{\hat{G}}_{\Pi }(t,\tau ) &=&\mathrm{diag}\{f_{0}(t,\tau
),f_{0}(t,\tau ),f_{\Pi }(t,\tau ),0\},\ \ \ \  \\
\mathbb{\hat{G}}_{\Phi }(t,\tau ) &=&\mathrm{diag}\{f_{\Phi }(t,\tau
),f_{\Phi }(t,\tau ),f_{0}(\tau ,t),0\},\ \ \ 
\end{eqnarray}
with the definitions 
\end{subequations}
\begin{subequations}
\label{desviaciones}
\begin{eqnarray}
f_{0}(t,\tau ) &=&e^{-\gamma _{\Phi }\tau }P_{\Pi }(t+\tau /2)-P_{\Pi
}(t)P_{\Phi }(\tau ), \\
f_{\Phi }(t,\tau ) &=&P_{\Phi }(t+\tau )-P_{\Phi }(t)P_{\Phi }(\tau ), \\
f_{\Pi }(t,\tau ) &=&P_{\Pi }(t+\tau )-P_{\Pi }(t)P_{\Pi }(\tau ).
\label{efePi}
\end{eqnarray}
These functions measure the transient departure from the validity of the
QRT. Only when the decay behaviors are exponential, they vanish identically
and the QRT is valid at all times. This situation happens when the evolution
is Markovian.

\subsection{Transient decay behaviors}

In order to illustrate the previous results, we specify the properties of
the complex environment, which in the context of the GBMA means to
characterize the set $\{\gamma _{R},P_{R}\}$. We choose 
\end{subequations}
\begin{equation}
\gamma _{R}=\gamma _{0}\exp [-bR],\;\;\;\;\;\;P_{R}=\frac{(1-e^{-a})}{%
(1-e^{-aN})}\exp [-aR],  \label{expo}
\end{equation}%
where $R\in \lbrack 0,N-1],$ $\gamma _{0}$ scale the random rates, and the
dimensionless constants $b$ and $a$ measure the exponential decay of the
random rates and their corresponding weights. The relevant parameters of
this set are%
\begin{equation}
\gamma \equiv \langle \gamma _{R}\rangle ,\ \ \ \ \ \ \ \ \beta \equiv \frac{%
\langle \gamma _{R}^{2}\rangle -\langle \gamma _{R}\rangle ^{2}}{\langle
\gamma _{R}\rangle },\ \ \ \ \ \alpha \equiv \frac{a}{b}.
\end{equation}
Here, $\gamma $ is the average rate\ and $\beta $ measures the dispersion of
the random rate set. On the other hand, in the limit $N\rightarrow \infty $
the set Eq.~(\ref{expo}) may leads to system dynamics characterized by a
power law behavior whose exponent is given by $\alpha $ \cite{gbma}.

\begin{figure}[tbph]
\includegraphics[width=9.cm]{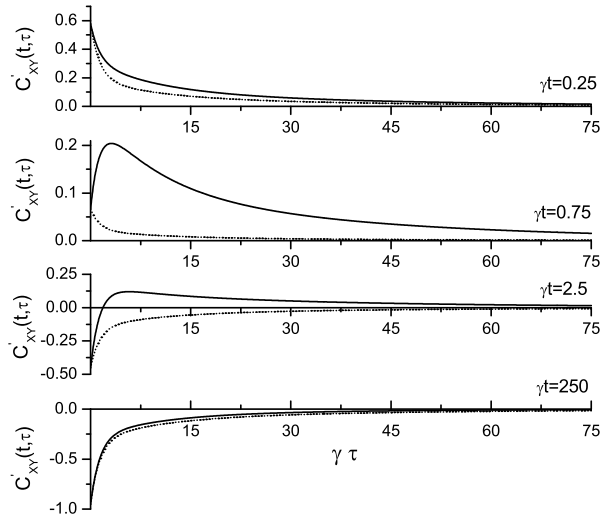}
\caption{Transient decay behavior of $C_{XY}^{\prime }(t,\protect\tau %
)\equiv C_{XY}(t,\protect\tau )/i.$ The parameters of the complex
environment are $b=2.15,$ $a=\protect\alpha b,$ $\protect\alpha =1/2$, $N=5$%
, $n_{th}=0,$ and $\protect\gamma _{\Phi }/\protect\gamma =0.02.$ The
dispersion rate results $\protect\beta /\protect\gamma =0.4.$ The dotted
lines correspond to the QRT. From top to bottom, we set $\protect\gamma %
t=0.25,$ $0.75,$ $2.5,$ and $250.$}
\label{Figure1QRT}
\end{figure}

In Fig.~\ref{Figure1QRT} we plot the transient decay behavior of the
correlation%
\begin{equation}
C_{XY}(t,\tau )\equiv \overline{\sigma _{x}(t)\sigma _{y}(t+\tau )},
\end{equation}%
which from Eq.~(\ref{diagonal}) and (\ref{Fextra}) can be written as%
\begin{equation}
C_{XY}(t,\tau )=i\{P_{\Phi }(\tau )S_{Z}(t)+f_{0}(t,\tau )[S_{Z}(0)-\mathcal{%
S}_{Z}^{\infty }]\},  \label{XY}
\end{equation}%
with $S_{Z}(t)=\mathcal{S}_{Z}^{\infty }+P_{\Pi }(t)[S_{Z}(0)-\mathcal{S}%
_{Z}^{\infty }].$\ We have chosen a cero temperature reservoir, $n_{th}=0$,
characterized by the random rate set Eq.~(\ref{expo}). As initial condition
we take the pure state $\left\vert +\right\rangle .$ Thus, $\mathcal{S}%
_{Z}^{\infty }=-1$ and $S_{Z}(0)=1.$ Notice that the initial value of each
plot describe the decay of the initial condition from the upper to the lower
state. In fact $C_{XY}(t,0)=iS_{Z}(t).$

As can be seen from the graphics, the predictions of the QRT are
asymptotically valid in the stationary regime, where the function $%
f_{0}(t,\tau )$ vanish identically. In fact, the correlation behavior
predicted by the QRT follows from Eq.~(\ref{XY}) after replacing $%
f_{0}(t,\tau )\rightarrow 0.$

The transient deviations from the QRT are proportional to the departure of
the system decay behavior from an exponential one. This departure arises
from the competence between the exponential decay introduced by the rate $%
\gamma _{\Phi }$ and the non-Markovian effects induced by the random rate
dispersion. From Eq.~(\ref{XY}) it is evident that the dispersive rate $%
\gamma _{\Phi }$\ introduces a global exponential decay. Thus, in general,
by increasing this rate, the transient deviation from the QRT are
diminished. On the other hand, an increasing of $\beta $ implies a strong
deviation from an exponential decay.

\begin{figure}[tbph]
\includegraphics[width=8.5cm]{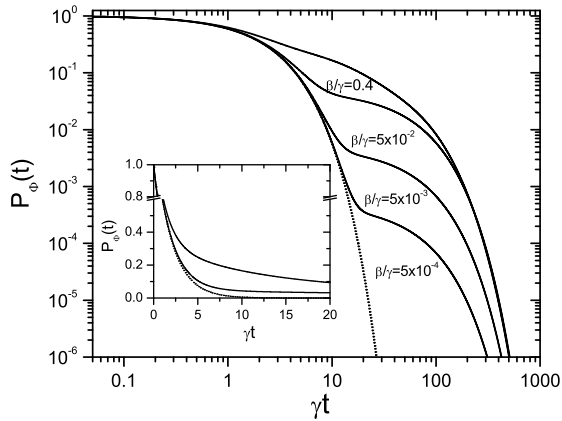}
\caption{Decay behavior of $P_{\Phi }(t).$ From top to bottom, the
parameters of the complex environment are $b=2.15$, $6.05$, $10.6$ and $%
15.2. $ In all cases we take $a=\protect\alpha b,$ $\protect\alpha =1/2$, $%
N=5$, $n_{th}=0,$ and $\protect\gamma _{\Phi }/\protect\gamma =0.02.$ The
dotted line corresponds to the Markovian decay $\exp [-t(\protect\gamma %
_{\Phi }+\protect\gamma /2)]$. The inset shows the same curves in a linear
plot. In this scale, the decay for $b=10.6$ and $15.2$ are indistinguishable
from the exponential one.}
\label{Figure2QRT}
\end{figure}

In order to enlighten the dependence in the dispersion of the random rate
set, in Fig.~\ref{Figure2QRT} we plot $P_{\Phi }(t)$, Eq.~(\ref{survivals}),
for different values of the dispersion rate $\beta $. This function
determine both the coherence decay \cite{footnote4} and the deviations $%
f_{0}(t,\tau )$ and $f_{\Phi }(t,\tau ),$ Eq.~(\ref{desviaciones}).

The short time behavior can be approximated by the exponential decay $%
P_{\Phi }(t)\simeq \exp [-\{\gamma _{\Phi }+\langle \gamma _{R}\rangle
/2\}t] $, while the asymptotic one by $P_{\Phi }(t)\simeq \exp [-\{\gamma
_{\Phi }+\langle 1/\gamma _{R}\rangle /2\}t]$. These behaviors can be
straightforwardly obtained from Eq.~(\ref{survivals}). In the intermediate
regime the decay is approximately a power law with exponent $\alpha $. By
diminishing $\beta $, the non-exponential decay behaviors occurs at small
values of $P_{\Phi }(t)$. In fact, for $\beta /\gamma \ll 1,$ the whole
decay may be well approximated by $P_{\Phi }(t)\simeq \exp [-\{\gamma _{\Phi
}+\langle \gamma _{R}\rangle /2\}t]$ (see inset). Thus, in this situation,
the QRT may be assumed valid\ at all times for correlations involving the
deviations $f_{0}(t,\tau )\approx 0$ and $f_{\Phi }(t,\tau )\approx 0.$ On
the other hand, as the population decay $P_{\Pi }(t)$ \cite{footnote4} does
not involves the dispersive rate $\gamma _{\Phi }$, in general we can not
disregard the transient effects introduced by $f_{\Pi }(t,\tau ),$ Eq.~(\ref%
{efePi}).

\subsection{Decay under the action of an external field}

For dealing with a manageable dynamics, we consider the external Hamiltonian 
$H_{ext}(t)=(\hbar \Omega /2)(\sigma ^{\dagger }e^{-i\omega _{A}t}+\sigma
e^{+i\omega _{A}t}).$ Then, the system density matrix dynamics can be
associated with a spin subject to a resonant external magnetic field \cite%
{blum} or with a two level optical transition driven by a resonant laser
field \cite{carmichael}.

In an interaction representation with respect to $\hbar \omega _{A}\sigma
_{z}/2$, the effective system Hamiltonian reads $H_{S}^{eff}=\hbar \Omega
\sigma _{x}/2.$ Thus, the evolution of the states $\rho _{R}(t)$ is given by
Eq.~(\ref{randommatrix}) with\ $H_{S}\rightarrow H_{S}^{eff}$ (see Appendix
A). From Eq.~(\ref{promedio}) and (\ref{kernel}), the expectation values of
the Pauli matrixes evolve as 
\begin{subequations}
\label{optical}
\begin{eqnarray}
\frac{dS_{X}(t)}{dt} &=&-\int_{0}^{t}d\tau \Gamma _{X}(t-\tau )S_{X}(\tau ),
\label{opticalX} \\
\frac{dS_{Y}(t)}{dt} &=&-\Omega S_{Z}(t)-\int_{0}^{t}d\tau \{\Gamma
_{Y}(t-\tau )S_{Y}(\tau )  \notag \\
&&+\Upsilon (t-\tau )[S_{Z}(\tau )-\mathcal{S}_{Z}^{\infty }]\}, \\
\frac{dS_{Z}(t)}{dt} &=&\Omega S_{Y}(t)+\int_{0}^{t}d\tau \{\Upsilon (t-\tau
)S_{Y}(\tau )  \notag \\
&&-\Gamma _{Z}(t-\tau )[S_{Z}(\tau )-\mathcal{S}_{Z}^{\infty }]\}.
\end{eqnarray}
In Appendix A we give the exact expressions for the kernels $\Gamma _{J}(t),$
$j=x,y,z,$ and $\Upsilon (t),$ as well as the expression for the non local
superoperator $\mathbb{L}(u),$ Eq.~(\ref{memory}). We remark that
independently of the set of random rates $\{\gamma _{R}\}$ and weights $%
\{P_{R}\}$, the kernels that define the evolution Eq.~(\ref{optical}) depend
explicitly on the intensity parameter $\Omega .$

The stationary state corresponding to the evolution of each state $\rho
_{R}(t)$, Eq.~(\ref{randommatrix}), reads 
\end{subequations}
\begin{equation}
\rho _{R}^{\infty }=\frac{1}{2}\left\{ \mathrm{I}+\frac{\Omega \gamma
_{R}\sigma _{y}-\gamma _{R}\gamma _{R}^{\Phi }\sigma _{z}}{%
(1+2n_{th})[\gamma _{R}\gamma _{R}^{\Phi }+\Omega ^{2}]}\right\} ,
\label{estacionaria}
\end{equation}
\begin{figure}[tbph]
\includegraphics[width=8.5cm]{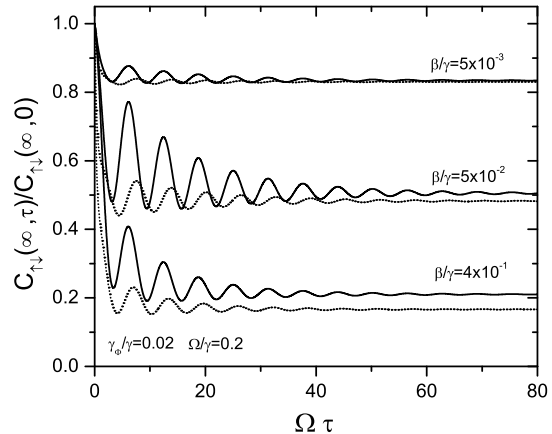}
\caption{Stationary decay behavior of $C_{\uparrow \downarrow }(t,\protect%
\tau ).$ From top to bottom, the parameters of the complex environment are $%
b=10.6$, $6.05$, and $2.15.$ In all cases we take $a=\protect\alpha b,$ $%
\protect\alpha =1/2$, $N=5$, $n_{th}=0,$ and $\protect\gamma _{\Phi }/%
\protect\gamma =0.02.$ The intensity is $\Omega /\protect\gamma =0.2.$ The
dotted lines correspond to the QRT.}
\label{Figure3QRT}
\end{figure}
which explicitly depends on\ $\gamma _{R}$ if $\Omega \neq 0$. Then, when
the system is subject to the action of the external field the QRT in not
fulfilled, even in the asymptotic regime. Consistently, as was demonstrated
in Ref.~\cite{carmichaelPRA}, the underlying Markovian evolution Eq.~(\ref%
{randommatrix}) does not satisfy the detailed balance condition Eq.~(\ref%
{condiciones}), neither the superoperator $\mathbb{L}(u)$ [Eq.~(\ref{L(u)})]
satisfy Eq.~(\ref{classic}). As the QRT is not fulfilled, the operators
correlations must to be calculated from the microscopic Hamiltonian
dynamics, which in our case implies the averaging procedure corresponding to
the GBMA.

In the next figures we characterize the correlation 
\begin{eqnarray}
C_{\uparrow \downarrow }(t,\tau ) &\equiv &\{C_{XX}(t,\tau )+C_{YY}(t,\tau
)\}/4 \\
&&-i\{C_{XY}(t,\tau )-C_{YX}(t,\tau )\}/4,  \notag
\end{eqnarray}%
where $C_{jk}(t,\tau )\equiv \overline{\sigma _{j}(t)\sigma _{k}(t+\tau )}$
are the correlations of the Pauli matrixes. Then it follows $C_{\uparrow
\downarrow }(t,\tau )=\overline{\sigma ^{\dagger }(t)\sigma (t+\tau )}$.
Each contribution $C_{jk}(t,\tau )$ can be determine from Eq.~(\ref%
{propagada}), which involves an average of the corresponding Markovian
solutions over the random rate set, Eq.~(\ref{expo}). On the other hand, the
QRT predictions follows from Eq.~(\ref{noMarkov}) with $\mathbf{F}(t,\tau
)\rightarrow 0.$

In Fig.~\ref{Figure3QRT} we plot the stationary decay $C_{\uparrow
\downarrow }(\infty ,\tau )/C_{\uparrow \downarrow }(\infty ,0)$, where $%
C_{\uparrow \downarrow }(\infty ,0)=[1+S_{Z}(\infty )]/2$, for different
values of the rate $\beta .$ We note that both the decay behaviors and
stationary values differ from the QRT predictions. As the evolution of $%
S_{X}(t)$ does not depend on $\Omega $ [see Eq.~(\ref{opticalX})], in the
asymptotic regime $C_{XX}(t,\tau )$ satisfies the QRT. Furthermore, as $%
\lim_{t\rightarrow \infty }S_{X}(t)=0,$ from Eq.~(\ref{coherente}) and (\ref%
{coherenteQRT}) we deduce that the disagreement in the asymptotic values
with respect to the QRT only arises due to the contribution $C_{YY}(t,\tau )$%
, while $C_{XY}(t,\tau )$ and $C_{YX}(t,\tau )$ only contribute to the
difference in the decay behaviors.

\begin{figure}[tbph]
\includegraphics[width=8.5cm]{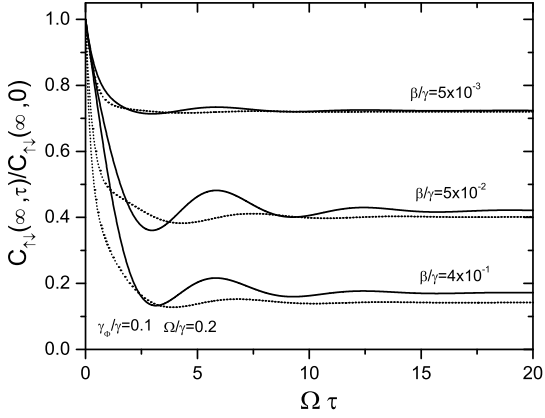}
\caption{Stationary decay behavior of $C_{\uparrow \downarrow }(t,\protect%
\tau )$ after increasing the dispersive rate, $\protect\gamma _{\Phi }/%
\protect\gamma =0.1$. The remaining parameters are the same than in Fig.~%
\protect\ref{Figure3QRT}. The dotted lines correspond to the decay predicted
by the QRT.}
\label{Figure4QRT}
\end{figure}

The asymptotic value predicted from Eq.~(\ref{coherente}) is%
\begin{equation}
C_{\uparrow \downarrow }(\infty ,\infty )=\frac{1}{(1+2n_{th})^{2}}%
\left\langle \Big(\frac{\Omega \gamma _{R}/2}{\gamma _{R}\gamma _{R}^{\Phi
}+\Omega ^{2}}\Big)^{2}\right\rangle ,
\end{equation}%
while from Eq.~(\ref{coherenteQRT}), for the QRT we get%
\begin{equation}
C_{\uparrow \downarrow }(\infty ,\infty )\overset{QRT}{=}\frac{1}{%
(1+2n_{th})^{2}}\left\langle \frac{\Omega \gamma _{R}/2}{\gamma _{R}\gamma
_{R}^{\Phi }+\Omega ^{2}}\right\rangle ^{2},
\end{equation}%
where we have used the stationary state Eq.~(\ref{estacionaria}). As can be
seen in the graphics, the difference between both predictions grows by
increasing the dispersion rate $\beta .$

In Fig.~\ref{Figure4QRT} , we plot the same correlation after increasing the
dispersive rate $\gamma _{\Phi }$ and maintaining fixed all other
parameters. We note that the deviations with respect to the QRT are
diminished. In fact, for small values of $\beta /\gamma $, the dynamical
deviations goes asymptotically to zero.

\begin{figure}[tbph]
\includegraphics[width=8.5cm]{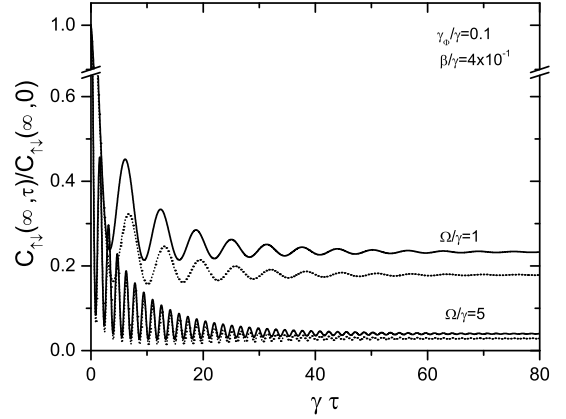}
\caption{Stationary decay behavior of $C_{\uparrow \downarrow }(t,\protect%
\tau )$ for different values of the intensity parameter. From top to bottom,
we take $\Omega /\protect\gamma =1$ and $5$. In both cases the parameters of
the complex environment are $b=2.15,$ $a=\protect\alpha b,$ $\protect\alpha %
=1/2$, $N=5$, $n_{th}=0,$ and $\protect\gamma _{\Phi }/\protect\gamma =0.1.$
The dotted lines correspond to the decay predicted by the QRT.}
\label{Figure5QRT}
\end{figure}

In Fig.~\ref{Figure5QRT} we plot $C_{\uparrow \downarrow }(\infty ,\tau )$
for different values of the field intensity $\Omega $. The deviations with
respect to the QRT are diminished by increasing $\Omega $. Even more, in the
limit of high intensity, the dynamical deviations vanish.%

The previous parameter dependence analysis relies in a specific correlation
and random rate set model. Similar conclusion can be obtained by studying
the asymptotic behaviors predicted by Eq.~(\ref{coherente}) and Eq.~(\ref%
{coherenteQRT}) for arbitrary correlations and random rate sets. The
deviations between both equations are proportional to the random dispersion
of the stationary state $\rho _{R}^{\infty },$\ Eq.~(\ref{estacionaria}).
Thus, as a measure of the departure from the validity of the QRT in the
stationary regime, we introduce the matrix%
\begin{equation}
\Xi _{JK}\equiv \big\langle S_{J}^{(R)}(\infty )S_{K}^{(R)}(\infty )%
\big\rangle-\big\langle S_{J}^{(R)}(\infty )\big\rangle\big\langle %
S_{K}^{(R)}(\infty )\big\rangle,
\end{equation}%
with $j,k=x,y,z,$ and where $S_{j}^{(R)}(\infty )\equiv \mathrm{Tr}%
_{S}\{\rho _{R}^{\infty }\sigma _{j}\}$ are the random stationary Pauli
expectation values. A general characterization of this matrix can be given
in a small and high intensity limits.

In the \textit{small intensity limit}, $\Omega \ll \{\gamma _{R}\},$ we can
approximate%
\begin{equation}
\Xi _{JK}\approx \frac{\Omega ^{2}}{(1+2n_{th})^{2}}\{\langle (\tau
_{R}^{\Phi })^{2}\rangle -\langle \tau _{R}^{\Phi }\rangle ^{2}\}+O(\Omega
^{3}),  \label{baja}
\end{equation}%
where $\tau _{R}^{\Phi }\equiv (\gamma _{R}/2+\gamma _{\Phi })^{-1}=1/\gamma
_{R}^{\Phi }.$ Consistently, $\Xi _{JK}$ goes to zero in the limit of small
intensity $\Omega $. On the other hand, by increasing the dispersive rate $%
\gamma _{\Phi }$, each contribution in Eq.~(\ref{baja}) diminish, which in
turn means that the predictions of the QRT approach the exact dynamics.

In the \textit{high intensity limit}, $\Omega \gg \{\gamma _{R}\}$ we get%
\begin{equation}
\Xi _{JK}\approx \frac{\Omega ^{-2}}{(1+2n_{th})^{2}}\{\langle \gamma
_{R}^{2}\rangle -\langle \gamma _{R}\rangle ^{2}\}+O(\Omega ^{-3}).
\end{equation}%
This expression implies that by increasing $\Omega ,$ the validity of the
QRT is asymptotically recuperated. This result is consistent with the fact
that at high intensity values \cite{carmichaelPRA} the underlying Markovian
dynamics for $\rho _{R}(t)$ satisfies the detailed balance condition Eq.~(%
\ref{condiciones}). In fact, in this limit the stationary states $\rho
_{R}^{\infty }$ [Eq.~(\ref{estacionaria})] can be approximated by $\rho
_{R}^{\infty }\approx \mathrm{I}/2$, which as expected does not depends on
the random rate.

\section{Summary and Conclusions}

In this paper we obtained the conditions under which a QRT can be assumed
valid for quantum non-Markovian master equations defined by Lindblad
superoperators with memory elements. In order to work on the base of a full
Hamiltonian description we deduced our results from a GBMA. This
approximation in a natural way leads to these kind of equations. In this
context, we demonstrated that operator correlations follow from a weighted
average of a set of Markovian solutions, each one characterized by a
different dissipative rate.

From our analysis, we deduced that a non-Markovian QRT can only be granted
in an stationary regime if the evolution satisfies a non-Markovian quantum
detailed balance condition [Eq.~(\ref{classic})], which in turn is related
with the time reversal symmetry of the stationary dissipative dynamics. When
this is not the case, the QRT is not fulfilled at any time, and in
consequence, the only way of calculating operators correlations is from the
corresponding microscopic dynamics.\ We remark that the impossibility of
formulating a non-Markovian quantum regression theorem outside a stationary
regime can be also demonstrated from general dynamical arguments (see
Appendix B).

In general, the departure from the predictions of the QRT not only implies
differences in the decay behaviors, but also in the asymptotic values of the
operators correlations. The magnitude of these deviations are proportional
to the departure of the system dynamics from a semigroup dynamical behavior,
i.e., an exponential one.

As an example of our results we worked out the dynamics of a two level
system subject to the action of an external coherent field and a complex
thermal environment whose action can be described in a GBMA. Without the
external field, the QRT is valid in an asymptotic regime. Consistently, the
non-Markovian quantum detailed balance condition is also satisfied. The
presence of the external field invalidates the QRT, even in the stationary
regime. Nevertheless, in the limit of high intensity, or when the effect of
a Markovian dispersive contribution is dominant, the QRT is asymptotically
reestablished to lowest order in the corresponding expansion parameters.

The present results provides an step forward in the understanding of
non-Markovian open quantum systems dynamics. In fact, we have found solid
physical criteria for the possibility of using a QRT for calculating
operators correlations when the system dynamics is described by a non-local
Lindblad evolution.

\section*{Acknowledgments}

Most of this work was done under a fellowship from Max Planck Institute for
the Physics of Complex Systems, Dresden, Germany. The author also tank
financial support from MCEyC of Spain through Juan de la Cierva program.

\appendix

\section{Density matrix evolution}

Here we characterize the evolution of the density matrix elements for the
example developed in Section V.

\subsection{Markovian evolution}

By denoting the matrix elements as%
\begin{equation}
\rho _{R}(t)=\left( 
\begin{array}{cc}
\Pi _{+}^{(R)}(t) & \Phi _{+}^{(R)}(t) \\ 
\Phi _{-}^{(R)}(t) & \Pi _{-}^{(R)}(t)%
\end{array}%
\right) ,
\end{equation}%
in an interaction representation with respect to $\hbar \omega _{A}\sigma
_{z}/2,$ from Eq.~(\ref{randommatrix}) we get the evolutions 
\begin{subequations}
\label{matrix}
\begin{eqnarray}
\frac{d}{dt}\Pi _{\pm }^{(R)}(t) &=&\gamma _{R}\{\mp \Pi _{-}^{eq}\Pi
_{+}^{(R)}(t)\pm \Pi _{+}^{eq}\Pi _{-}^{(R)}(t)\} \\
&&\pm \frac{i\Omega }{2}[\Phi _{+}^{(R)}(t)-\Phi _{-}^{(R)}(t)],  \notag \\
\frac{d}{dt}\Phi _{\pm }^{(R)}(t) &=&\pm \frac{i\Omega }{2}[\Pi
_{+}^{(R)}(t)-\Pi _{-}^{(R)}(t)]-\gamma _{R}^{\Phi }\Phi _{\pm }^{(R)}(t).\
\ \ \ \ \ \ \ 
\end{eqnarray}
The operators expectation values are defined by the evolutions 
\end{subequations}
\begin{subequations}
\label{opticalMarkov}
\begin{eqnarray}
\frac{dS_{X}^{(R)}(t)}{dt} &=&-\gamma _{R}^{\Phi }S_{X}^{(R)}(t), \\
\frac{dS_{Y}^{(R)}(t)}{dt} &=&-\Omega S_{Z}^{(R)}(t)-\gamma _{R}^{\Phi
}S_{Y}^{(R)}(t), \\
\frac{dS_{Z}^{(R)}(t)}{dt} &=&\Omega S_{Y}^{(R)}(t)-\gamma
_{R}[S_{Z}^{(R)}(t)-\mathcal{S}_{Z}^{\infty }],
\end{eqnarray}
where $S_{j}^{(R)}(t)\equiv \mathrm{Tr}_{S}\{\rho _{R}(t)\sigma _{j}\}$,
with $\sigma _{j}$ the Pauli matrixes.

\subsection{Non-Markovian evolution}

Here, for arbitrary set $\{\gamma _{R},P_{R}\},$ we present the exact
expressions for the kernels that define the evolution Eq.~(\ref{optical}).
From Eq.~(\ref{kernel}) and Eq.~(\ref{opticalMarkov}), we get 
\end{subequations}
\begin{subequations}
\begin{eqnarray}
\Gamma _{X}(u) &=&K_{\Phi }(u), \\
\Gamma _{Y}(u) &=&D\{(u+C)[\frac{B}{2}+(u+\gamma _{\Phi })]+\Omega
^{2}\}+\gamma _{\Phi },\ \ \ \ \ \ \  \\
\Gamma _{Z}(u) &=&2D\{(u+B)[\frac{C}{2}+(u+\gamma _{\Phi })]+\Omega ^{2}\},
\\
\Upsilon (u) &=&D(C-B)\Omega ,
\end{eqnarray}
where $D$ denotes the function 
\end{subequations}
\begin{equation}
D(u)=\frac{B(u)/2}{[u+B(u)][u+B(u)/2+\gamma _{\Phi }]+\Omega ^{2}}.
\end{equation}%
The extra function $B$ and $C$ are defined by%
\begin{equation}
B(u)=\frac{\langle T(u)\gamma _{R}\rangle }{\langle T(u)\rangle },\ \ \ \ \
\ \ \ \ C(u)=\frac{\langle T(u)(\gamma _{R})^{2}\rangle }{\langle T(u)\gamma
_{R}\rangle }
\end{equation}%
where we have introduced 
\begin{equation}
T(u)=\frac{1/2}{(u+\gamma _{R})[u+\gamma _{R}/2+\gamma _{\Phi }]+\Omega ^{2}}%
.
\end{equation}
The corresponding density matrix evolution can be written as 
\begin{subequations}
\label{density}
\begin{eqnarray}
\dot{\Pi}_{\pm }(u) &=&\Gamma _{Z}(u)\{\mp \Pi _{-}^{eq}\Pi _{+}(u)\pm \Pi
_{+}^{eq}\Pi _{-}(u)\},  \notag \\
&&\pm \frac{i\Omega (u)}{2}[\Phi _{+}(u)-\Phi _{-}(u)], \\
\dot{\Phi}_{\pm }(u) &=&\pm \frac{i\Omega (u)}{2}[\Pi _{+}(u)-\Pi
_{-}(u)]\pm i\frac{\Upsilon (u)/(2u)}{1+2n_{th}}  \notag \\
&&-\Gamma _{\Phi }^{+}(u)\Phi _{\pm }(u)-\Gamma _{\Phi }^{-}(u)\Phi _{\mp
}(u),
\end{eqnarray}
where, for shortening the notation, $\dot{g}(u)\equiv ug(u)-g(0)$\ denotes
the Laplace transform of the time derivative of an arbitrary function $g(t)$%
. In the inhomogeneous term for the coherences, we have used $\Pi
_{+}(u)+\Pi _{-}(u)=1/u$. Furthermore, we have defined $\Omega (u)\equiv
\Omega +\Upsilon (u)$ and $\Gamma _{\Phi }^{\pm }(u)\equiv \frac{1}{2}%
[\Gamma _{X}(u)\pm \Gamma _{Y}(u)].$ The stationary state reads 
\end{subequations}
\begin{equation}
\rho _{S}^{\infty }=\frac{1}{2}\left\{ \mathrm{I}+\frac{\Omega \Gamma
_{Z}\sigma _{y}-[\Gamma _{Y}\Gamma _{Z}+\Upsilon (\Upsilon +\Omega )]\sigma
_{z}}{(1+2n_{th})[\Gamma _{Y}\Gamma _{Z}+(\Upsilon +\Omega )^{2}]}\right\} ,
\end{equation}%
with the notation $\Gamma _{J}\equiv \left. \Gamma _{J}(u)\right\vert _{u=0}$%
. Consistently, after some algebra, it is possible to write this state as an
average of the corresponding Markovian stationary states, i.e., $\rho
_{S}^{\infty }=\langle \rho _{R}^{\infty }\rangle $, where $\rho
_{R}^{\infty }$ is defined by Eq.~(\ref{estacionaria}).

The superoperator $\mathbb{L}(u)$ [Eq.~(\ref{memory})] corresponding to the
evolution Eq.~(\ref{density}) can be written as non-diagonal non-local
Lindblad superoperator%
\begin{equation}
\mathbb{L}(u)[\bullet ]=\mathcal{L}_{H}(u)[\bullet ]+\frac{1}{2}\sum_{\alpha
\beta }a_{\alpha \beta }(u)([V_{\alpha },\bullet V_{\beta }^{\dagger
}]+[V_{\alpha }\bullet ,V_{\beta }^{\dagger }]),  \label{L(u)}
\end{equation}%
with the operators $\{V_{\alpha }\}_{\alpha =1,2,3}=\{\sigma ,\sigma
^{\dagger },\sigma _{z}\}$. The Hamiltonian contribution reads%
\begin{equation}
\mathcal{L}_{H}(u)[\bullet ]=-i\frac{\Upsilon (u)}{2}[\sigma _{x},\bullet ],
\end{equation}%
and the matrix elements $a_{\alpha \beta }(u)$ are defined by%
\begin{eqnarray}
a_{11}(u) &=&\Pi _{-}^{eq}\Gamma _{Z}(u), \\
a_{22}(u) &=&\Pi _{+}^{eq}\Gamma _{Z}(u), \\
a_{33}(u) &=&\frac{1}{4}\{\Gamma _{X}(u)+\Gamma _{Y}(u)-\Gamma _{Z}(u)\}, \\
a_{12}(u) &=&a_{21}(u)=-\frac{1}{2}\{\Gamma _{X}(u)-\Gamma _{Y}(u)\}, \\
a_{13}(u) &=&a_{23}(u)=-i\frac{\Upsilon (u)}{4(1+2n_{th})}, \\
a_{31}(u) &=&a_{32}(u)=i\frac{\Upsilon (u)}{4(1+2n_{th})}.
\end{eqnarray}

Without the external excitation, $\Omega =0$, the superoperator $\mathbb{L}%
(u)$ reduce to Eq.~(\ref{lindbladU}). Furthermore, when the coherence decay
behavior can be approximated by an exponential one, $P_{\Phi }(t)=e^{-\gamma
_{\Phi }t}P_{\Pi }(t/2)\simeq \exp [-(\gamma _{\Phi }+\langle \gamma
_{R}\rangle /2)t]$, the density matrix evolution can be written in a Schr%
\"{o}dinger representation as%
\begin{eqnarray}
\frac{d\rho _{S}(t)}{dt} &=&-\frac{i}{\hbar }[H_{S},\rho _{S}(t)]+\frac{%
\gamma _{\Phi }}{2}\mathcal{L}_{\Phi }[\rho _{S}(t)]  \label{Master} \\
&&+\frac{1}{1+2n_{th}}\int_{0}^{t}d\tau K(t-\tau )\mathcal{L}_{th}[\rho
_{S}(\tau )],  \notag
\end{eqnarray}%
with $H_{S}=\hbar \omega _{A}\sigma _{z}/2$. This expression relies in the
validity of the approximation $K(u\pm iw_{A})\simeq K(\infty )=\langle
\gamma _{R}\rangle $, which can be considered always valid if $w_{A}$ is an
optical frequency. Furthermore, if $\gamma _{\Phi }\ll \langle \gamma
_{R}\rangle $ the dispersive contribution can be drop. In general this last
condition is valid when the decay of $P_{\Pi }(t)$ develops two strong
different time scales. For example, consider a random rate\ that assumes
only two different values $\gamma _{\uparrow /\downarrow }$, with
probabilities $P_{\uparrow /\downarrow }$. Then $P_{\Pi }(t)=P_{\uparrow
}e^{-\gamma _{\uparrow }t}+P_{\downarrow }e^{-\gamma _{\downarrow }t}$.
Under the conditions $P_{\downarrow }\ll P_{\uparrow }$ and $\gamma
_{\downarrow }\ll \gamma _{\Phi }\ll \gamma _{\uparrow }$, we can
approximate $P_{\Pi }(t)e^{-\gamma _{\Phi }t}\approx e^{-\left\langle \gamma
_{R}\right\rangle t}+O(P_{\downarrow }/P_{\uparrow }).$ Another examples
follow from the decay of Fig.~\ref{Figure2QRT} for small $\beta /\gamma .$
On the other hand, Eq.~(\ref{Master}) can also be assumed valid in presence
of the external field if the exact kernels are taken to cero order in the
intensity parameter $\Omega .$

\section{On the impossibility of formulating a non-Markovian quantum
regression theorem at all times}

The impossibility of formulating a non-Markovian regression theorem outside
a stationary regime can be demonstrated on general dynamical arguments. In
fact, it is simple to proof that the validity of the quantum regression
theorem at all times is \textit{only} compatible with a Markovian dynamics.
This affirmation seems to contradict our main conclusions. Nevertheless,
here we demonstrate that this result confirm the correctness of our approach.

First, we write the system density matrix as 
\begin{equation}
\rho _{S}(t)=\mathbb{T}(t)[\rho _{S}(0)],  \label{dynamics}
\end{equation}%
where $\mathbb{T}(t)$ is the propagator corresponding to the evolution Eq.~(%
\ref{noMarkovMaster}). Then, it is defined in the Laplace domain by%
\begin{equation}
\mathbb{T}(u)=\frac{1}{u-[\mathcal{L}_{H}+\mathbb{L}(u)]}.
\end{equation}%
In terms of this object, we can write the operator expectation values as 
\begin{subequations}
\begin{eqnarray}
\overline{\mathbf{A}(t)} &=&\mathrm{Tr}_{S}\{\mathbf{A}(0)\mathbb{T}(t)[\rho
_{S}(0)]\},  \label{Aver} \\
&=&\mathrm{Tr}_{S}\{\rho _{S}(0)\mathbb{T}^{\#}(t)[\mathbf{A}(0)]\},
\end{eqnarray}%
where the second line defines the dual propagator $\mathbb{T}^{\#}(t).$ By
assuming valid the quantum regression theorem, the operator correlations can
be written as \cite{carmichael,cohen,loudon,lax} 
\end{subequations}
\begin{equation}
\overline{O(t)\mathbf{A}(t+\tau )}=\mathrm{Tr}_{S}\{\rho _{S}(t)O(0)\mathbb{T%
}^{\#}(\tau )[\mathbf{A}(0)]\}.
\end{equation}%
This expression must to be valid for arbitrary operators $O$ and $\mathbf{A}%
. $ In particular, by taking $O=I_{S},$ where $I_{S}$ is the system identity
operator, it follows 
\begin{subequations}
\begin{eqnarray}
\overline{\mathbf{A}(t+\tau )} &=&\mathrm{Tr}_{S}\{\rho _{S}(t)\mathbb{T}%
^{\#}(\tau )[\mathbf{A}(0)]\}, \\
&=&\mathrm{Tr}_{S}\{\mathbf{A}(0)\mathbb{T}(\tau )[\rho _{S}(t)]\}, \\
&=&\mathrm{Tr}_{S}\{\mathbf{A}(0)\mathbb{T}(\tau )\mathbb{T}(t)[\rho
_{S}(0)]\}.  \label{AverageCorrelation}
\end{eqnarray}%
On the other hand, from Eq.~(\ref{Aver}), we can write 
\end{subequations}
\begin{equation}
\overline{\mathbf{A}(t+\tau )}=\mathrm{Tr}_{S}\{\mathbf{A}(0)\mathbb{T}%
(t+\tau )[\rho _{S}(0)]\}.  \label{ATTau}
\end{equation}%
As $\mathbf{A}(0)$ is an arbitrary operator, by comparing this expression
and Eq.~(\ref{AverageCorrelation}), it follows%
\begin{equation}
\mathbb{T}(t+\tau )\rho _{S}(0)=\mathbb{T}(\tau )\mathbb{T}(t)\rho _{S}(0).
\label{Semigroup}
\end{equation}%
For arbitrary time $t<\infty $, and $\rho _{S}(0)\neq \rho _{S}^{\infty },$
where $\rho _{S}^{\infty }$\ is the stationary state corresponding to the
dynamics Eq.~(\ref{dynamics}), this equality can only be satisfied if the
propagator $\mathbb{T}(t)$ corresponds to a semigroup structure, i.e., a 
\textit{Markovian evolution}. Therefore, a regression theorem can be
satisfied at all times only when the dynamics does not has any memory
contribution. We notice that this result is in perfect agreement with our
main conclusions. In fact, we have found that a non-Markovian quantum
regression theorem may be valid (or not) \textit{only in a stationary regime}%
. In this limit, the previous calculations steps \textit{does not impose}
any constraint on the propagator $\mathbb{T}(t).$ This affirmation follows 
\textit{trivially} by taking $\rho _{S}(0)=\rho _{S}^{\infty }$ in Eq.~(\ref%
{Semigroup}), or equivalently by introducing the limit $t\rightarrow \infty
, $ 
\begin{equation}
\lim_{t\rightarrow \infty }\mathbb{T}(t+\tau )\rho
_{S}(0)=\lim_{t\rightarrow \infty }\mathbb{T}(\tau )\mathbb{T}(t)\rho
_{S}(0),
\end{equation}%
which, independently of the properties of $\mathbb{T}(t),$ deliver $\rho
_{S}^{\infty }=\rho _{S}^{\infty }.$ This last equality follows immediately
from $\mathbb{T}(\tau )[\rho _{S}^{\infty }]=\rho _{S}^{\infty },$
expression valid for any time $\tau .$ Alternatively, one can take the limit 
$t\rightarrow \infty $ in Eq.~(\ref{AverageCorrelation}) 
\begin{subequations}
\begin{eqnarray}
\lim_{t\rightarrow \infty }\overline{\mathbf{A}(t+\tau )}\! &=&\!\mathrm{Tr}%
_{S}\{\mathbf{A}(0)\mathbb{T}(\tau )\!\lim_{t\rightarrow \infty }\!\mathbb{T}%
(t)[\rho _{S}(0)]\},\ \ \ \ \ \  \\
\! &=&\!\mathrm{Tr}_{S}\{\mathbf{A}(0)\mathbb{T}(\tau )[\rho _{S}^{\infty
}]\}, \\
\! &=&\!\mathrm{Tr}_{S}\{\mathbf{A}(0)\rho _{S}^{\infty }\}.
\end{eqnarray}%
On the other hand, in the same limit, from Eq.~(\ref{ATTau}), as expected,
we get the same result 
\end{subequations}
\begin{equation}
\lim_{t\rightarrow \infty }\overline{\mathbf{A}(t+\tau )}=\mathrm{Tr}_{S}\{%
\mathbf{A}(0)\rho _{S}^{\infty }\}.
\end{equation}%
Therefore, the calculations steps that lead to the constraint Eq.~(\ref%
{Semigroup}) only contradict the possibility of establishing a non-Markovian
quantum regression theorem outside the stationary regime. These arguments
provide an alternative demonstration of the consistency and correctness of
our results.


\begin{thebibliography}{99}
\bibitem{artificial} In Ref.~\cite{jpa} a similar conclusion was obtained
from an abstract tripartite system-environment interaction. Here we proved
this result from an usual bipartite interaction.
\bibitem{footnote1} If the set $\{g_{nm}\}$ does not depends on the
parameter $\varepsilon $, by assuming valid the detailed balance condition
Eq.~(\ref{clasica}), one can find a continuous parametrized infinite family
of stationary solutions. In fact, in such a case, the detailed balance
condition can be written as $g_{nm}\sum_{k=0}^{\infty
}p_{m}^{(k)}(\varepsilon -\varepsilon _{0})^{k}/k!=g_{mn}\sum_{k=0}^{\infty
}p_{n}^{(k)}(\varepsilon -\varepsilon _{0})^{k}/k!,$ where $%
p_{n}^{(k)}\equiv (\partial ^{k}/\partial \varepsilon ^{k})p_{n}(\infty
,\varepsilon )|_{\varepsilon =\varepsilon _{0}}.$ As these relations must to
be valid for any $\varepsilon ,$ it follows that for any $k$ and $%
\varepsilon _{0}$ the set of derivatives $\{p_{n}^{(k)}\}$ allows to
construct a new and different stationary solution.
\bibitem{footnote2} Note that in terms of dual superoperators the
correlations Eqs.~(\ref{QRT1}) and (\ref{QRT2}), adopt a simple structure $%
\overline{O(t)\mathbf{A}(t+\tau )}=\langle \mathrm{Tr}_{S}\{\rho
_{R}(t)O(0)e^{(\mathcal{L}_{H}^{\#}+\mathcal{L}_{R}^{\#})\tau }[\mathbf{A}%
(0)]\}\rangle ,$ while for the correlation of three operators we get $%
\overline{O_{1}(t)\mathbf{A}(t+\tau )O_{2}(t)}=\langle \mathrm{Tr}_{S}\{\rho
_{R}(t)O_{1}e^{(\mathcal{L}_{H}^{\#}+\mathcal{L}_{R}^{\#})\tau }[\mathbf{A}%
]O_{2}\}\rangle .$
\bibitem{footnote3} By writing the Lindblad superoperator Eq.~(\ref{lindblad}%
) in the form $\mathcal{L}[\bullet ]=\sum_{\alpha \beta }a_{\alpha \beta
}(V_{\alpha }\bullet V_{\beta }^{\dagger }-\frac{1}{2}\{V_{\beta }^{\dagger
}V_{\alpha },\bullet \}_{+}),$ the dual superoperator $\mathcal{L}^{\#}$ can
be immediately expressed as $\mathcal{L}^{\#}[\bullet ]=\sum_{\alpha \beta
}a_{\alpha \beta }(V_{\beta }^{\dagger }\bullet V_{\alpha }-\frac{1}{2}%
\{V_{\beta }^{\dagger }V_{\alpha },\bullet \}_{+}),$ where $\{\cdot ,\cdot
\}_{+}$ denotes an anticonmutator operation. For the Hamiltonian
contribution if follows $\mathcal{L}_{S}^{\#}=-\mathcal{L}_{S}.$ On the
other hand, $\widetilde{\mathcal{L}}_{H}=-\mathcal{L}_{H},$ and when $%
\widetilde{V}_{\alpha }=V_{\alpha }^{\dagger }$, we get $\widetilde{\mathcal{%
L}}=\mathcal{L}.$
\bibitem{footnote3bis} As in Eq.~(\ref{condiciones}), the condition Eq.~(\ref%
{classic2}) can be split in two equations, corresponding respectively to the
Hamiltonian and dissipative contributions. Nevertheless, in the
non-Markovian case we must also to consider possible non-local Hamiltonian
contributions introduced by $\mathbb{L}(u)$ [see for example Eq.~(\ref{L(u)}%
)].
\bibitem{footnote4} The solutions of Eq.~(\ref{popular}) and (\ref%
{coherencia}) can be written as $\Pi _{\pm }(t)=P_{\Pi }(t)\Pi _{\pm
}(0)+[1-P_{\Pi }(t)]\Pi _{\pm }^{eq},$ while for the coherences it reads $%
\Phi _{\pm }(t)=P_{\Phi }(t)\Phi _{\pm }(0).$ From these expressions, it is
possible to write the system density matrix as $\rho _{S}(t)=P_{\Pi }(t)\rho
_{S}^{+}(0)+P_{\Phi }(t)\rho _{S}^{-}(0)+[1-P_{\Pi }(t)]\rho _{S}^{\infty }.$
This final expression is used when deriving Eq.~(\ref{Fextra}).
\end{thebibliography}
\end{document}